\pdfoutput=1
\RequirePackage{ifpdf}
\ifpdf 
\documentclass[pdftex]{sigma}
\else
\documentclass{sigma}
\fi

\newcommand{\x}{{\bf x}}
\newcommand{\D}{{\rm d}}

\numberwithin{equation}{section}

\begin{document}

\allowdisplaybreaks

\renewcommand{\PaperNumber}{057}

\FirstPageHeading

\ShortArticleName{Contractions of Quantum Superintegrable Systems and the Askey Scheme}

\ArticleName{Contractions of 2D 2nd Order Quantum
\\
Superintegrable Systems and the Askey Scheme
\\
for Hypergeometric Orthogonal Polynomials}

\Author{Ernest G.~KALNINS~$^\dag$, Willard MILLER Jr.~$^\ddag$ and Sarah POST~$^\S$}

\AuthorNameForHeading{E.G.~Kalnins, W.~Miller Jr.\ and S.~Post}

\Address{$^\dag$ Department of Mathematics, University of Waikato, Hamilton, New Zealand}
\EmailD{\href{mailto:math0236@waikato.ac.nz}{math0236@waikato.ac.nz}}

\Address{$^\ddag$ School of Mathematics, University of Minnesota, Minneapolis, MN, 55455, USA}
\EmailD{\href{mailto:miller@ima.umn.edu}{miller@ima.umn.edu}}

\Address{$^\S$ Department of Mathematics, U.~Hawai`i at Manoa, Honolulu, HI, 96822, USA}
\EmailD{\href{mailto:spost@hawaii.edu}{spost@hawaii.edu}}

\ArticleDates{Received May 29, 2013, in f\/inal form September 26, 2013; Published online October 02, 2013}

\Abstract{We show explicitly that all 2nd order superintegrable systems in 2 dimensions are limiting cases
of a~single system: the generic 3-parameter potential on the 2-sphere, ${\rm S}9$ in our listing.
We extend the Wigner--In\"on\"u method of Lie algebra contractions to contractions of quadratic algebras
and show that all of the quadratic symmetry algebras of these systems are contractions of that of ${\rm S}9$.
Amazingly, all of the relevant contractions of these superintegrable systems on f\/lat space and the sphere
are uniquely induced by the well known Lie algebra contractions of~e(2) and~so(3).
By contracting function space realizations of irreducible representations of the ${\rm S}9$ algebra (which give the
structure equations for Racah/Wilson polynomials) to the other superintegrable systems, and using Wigner's
idea of ``saving'' a~representation, we obtain the full Askey scheme of hypergeometric orthogonal
polynomials.
This relationship directly ties the polynomials and their structure equations to physical phenomena.
It is more general because it applies to all special functions that arise from these systems via separation
of variables, not just those of hypergeometric type, and it extends to higher dimensions.}

\Keywords{Askey scheme; hypergeometric orthogonal polynomials; quadratic algebras} \Classification{33C45;
33D45; 33D80; 81R05; 81R12}

\section{Introduction}

\looseness=1
A quantum superintegrable system is an integrable Hamiltonian system on an $n$-dimensional
Riemannian/pseudo-Riemannian manifold with potential, $H=\Delta_n+V$, that admits $2n-1$ algebraically
independent partial dif\/ferential operators commuting with~$H$, apparently the maximum possible.
Superintegrability captures the properties of quantum Hamiltonian systems that allow the Schr\"odinger
eigenvalue problem $H\Psi=E\Psi$ to be solved exactly, analytically and algebraically, see~\cite{MPWreview}
and references therein.
A system is of order $N$ if the maximum order of the symmetry operators, other than~$H$, is $N$.
The simplest examples of such systems are the 1st order superintegrable systems given by the potential-free
Hamiltonian on 2D Euclidean or Minkowski space and on the 2-sphere.
These are the Euclidean Helmholtz equation $(P_1^2+P_2^2)\Phi=-\lambda^2 \Phi$ (or the Klein--Gordon
equation $(P_1^2-P_2^2)\Phi=-\lambda^2 \Phi$), and the Laplace--Beltrami eigenvalue equation on the
2-sphere $(J_1^2+J_2^2+J_3^2)\Psi=-j(j+1)\Psi$. Here the symmetry algebras are Lie algebras.
The symmetry generators in the f\/irst case are $P_1$, $P_2$, $J$ and the commutators close at 1st order to form
the Lie algebra~e(2).
In the second case the generators are~$J_1$,~$J_2$,~$J_3$ and the symmetry algebra is the Lie algebra~so(3).
The irreducible representations of~so(3) are labeled by the integer $j$ and are $2j+1$ dimensional.
From this, one can deduce that the Hilbert solution space of the eigenvalue equation breaks into a~direct
sum of eigenspaces, each with eigenvalue $-j(j+1)$ for integer $j$ and with multiplicity $2j+1$.
One can f\/ind 2-variable dif\/ferential operator models of these irreducible representations in which the
eigenfunctions of~$J_3$ are the spherical har\-mo\-nics~$Y_{j,n}$.
Similarly we can f\/ind models of the inf\/inite dimensional irreducible representations of~e(2) for any
$\lambda>0$ such that the eigenfunctions of~$J$ are Bessel functions~$J_n$.

\looseness=1
It was exactly these systems which motivated the pioneering work of In\"on\"u and Wigner~\cite{Wigner} on
Lie algebra contractions.
While that paper introduced Lie algebra contractions in ge\-neral, the motivation and virtually all the
examples were of symmetry groups of these systems.
In~\cite{Wigner} it was shown that~so(3) contracts to~e(2).
In the physical space this can be accomplished by letting the radius of the sphere go to inf\/inity, so
that the surface f\/lattens out.
Under this limit the Laplace--Beltrami eigenvalue equation goes to the Helmholtz equation.
Also, the irreducible representations of the eigenspaces of the equation on the sphere go to those in
Euclidean space but only if one ``saves'' the representation by passing through a~sequence of values of $j$
going to inf\/inity~\cite{Wigner, Talman}.
Similarly the $2j+1$ dimensional models for so(3) go to models for e(2) as $j\to \infty$ and the spherical
harmonics converge to Bessel functions.
The various special functions that arise from these eigenvalue equations via separation of variables are
also related by this contraction, e.g.~\cite{IPSW, KMPog1999, Talman}.
In his lecture notes~\cite{Talman}, Wigner employed 1-variable dif\/ferential operator models of the
irreducible e(2) and so(3) representations.
In these cases, the irreducible representations are given by elementary functions, exponentials and
monomials, but they made it easy for Wigner to ``save'' a~representation for contractions and, as he pointed
out, they also told us the expansion coef\/f\/icients expressing one basis of separable solutions of the
original eigenvalue equations in terms of another, e.g.\ plane waves in terms of spherical waves.
This last is extended in the book~\cite{Miller} where the models are used to expand, say, parabolic
cylindrical wave solutions in terms of spherical solutions.

\looseness=1
In this paper we have extended these ideas to the more complicated case of 2nd order superintegrable
systems, still in 2D, where there are potentials.
All such systems are known.
There are about 58 types on a~variety of manifolds but under the St\"ackel transform~\cite{KKM20042}, an
invertible structure preserving mapping, they divide into 12 equivalence classes~\cite{KKMP2009, Kress2007}.
Now the symmetry algebra is a~quadratic algebra, not usually a~Lie algebra, and the irreducible
representations of this algebra determine the eigenvalues of~$H$ and their
multip\-li\-city~\cite{BDK,Dask2001,DASK2007,DASK2005,Zhedanov1992a,Zhedanov1992b,VILE,Zhedanov1991}.

We introduce the notion of quadratic algebra contractions in general, but focus on the special case of
contractions of superintegrable systems.
We demonstrate explicitly that up to St\"ackel transform, all the 2nd order superintegrable systems are
limiting cases of a~single system: the generic 3-parameter potential on the 2-sphere, ${\rm S}9$ in our listing.
Analogously all quadratic symmetry algebras of these systems are contractions of ${\rm S}9$.
Amazingly, all of the required quadratic algebra contractions are uniquely induced by Lie algebra
contractions of e(2) and so(3), the broken symmetries of the underlying spaces.
These contractions have been long since classif\/ied, e.g.\
\cite{WW}.
In this paper we just list the coordinate contractions and the induced limits of the potentials.
In a~forthcoming article~\cite{KMSH} the proof will be given that these Lie algebra contractions of so(3)
and e(2) uniquely lift to contractions of superintegrable systems on the sphere and f\/lat space, including
the potentials (modulo the choice of basis for the potentials).
Thus, the limits of the physical systems and associated limit of representations, while apparently somewhat
arbitrary, are in fact completely determined by the possible contractions of the algebras.
Again the eigenvalues of the Schr\"odinger operator can be computed from the irreducible representations of
the quadratic algebras and the multiplicities of the eigenvalues from the dimensions of these
representations.
Just as before we can f\/ind contractions that relate the physical systems and
we can ``save'' a~representation in the contraction of the quadratic algebras.

A new feature is that 1-variable dif\/ference operator models of the quadratic algebras become important.
Their eigenfunctions are special functions dif\/ferent from the separated solutions of the original quantum
operators and not the completely elementary functions of the 1st order superintegrable case.
Indeed, the irreducible representations of ${\rm S}9$ have a~realization in terms of dif\/ference operators in 1
variable, exactly the structure algebra for the Wilson and Racah polynomials~\cite{KMPost1}!
Indeed this algebra is exactly the Askey--Wilson algebra for $q=1$
and the Racah algebra QR(3)~\cite{GWH, granovskii1992mutual, Zhedanov1992a, Terwilliger}.
In a~recent paper, Genest, Vinet and Zhedanov~\cite{genest2013superintegrability} give an elegant,
algebraic proof of the equivalence between the symmetry algebra for ${\rm S}9$ and the Racah problem of
$\mathfrak{su}(1,1)$.
By contracting these representations to obtain the representations of the quadratic symmetry algebras of
the other superintegrable systems, we obtain the full Askey scheme of orthogonal hypergeometric
polynomials~\cite{KLS,Koorn}.
Thus under contractions we can relate eigenfunctions of the dif\/ference operator models as well as
separable solutions of the original Schr\"odinger eigenvalue equations.
The whole procedure is very natural and it is clear that the Askey scheme is directly related to the
contraction picture.

\looseness=-1
This relationship ties the structure equations directly to physical phenomena.
In some cases, e.g.\ for Hahn polynomials, we have contractions leading to structure equations not obeyed
by the full family of Hahn polynomials.
These families with higher than usual symmetry we refer to as ``special''.
The structure theory exposes these ``special'' systems in a~natural manner.
Similarly, for the Meixner--Pollaczek and Pseudo Jacobi polynomials and special cases of the Wilson,
continuous Hahn and continuous dual Hahn polynomials, there are instances where the parameters in these
functions occur in complex conjugate pairs and a~real three term recurrence relation exists, so that the
polynomials are orthogonal with respect to a~positive weight function.
In these instances, the quantum Hamiltonian is {\it PT-symmetric} in the sense that the scalar potential
$V(x,y)$ satisf\/ies $V(x,y)=\overline{V(y,x)}$.
Here $T$ means time inversion (complex conjugation) and $P$ means permutation.
When the Hamiltonian admits $PT$ symmetry then even though the potential is complex, the bound state energy
eigenvalues must be real.
{}$PT$ symmetry in physics is controversial but the mathematics is clear~\cite{Mos2002}.
The usual meaning of $P$ in physics is space inversion, so that $PT$ symmetry requires
$V(x,y)=\overline{V(-x,-y)}$, but the outcome is the same: the potential is complex but the energy
eigenvalues are real.
The superintegrable system $S6$, an analog of the 2D hydrogen atom on the 2-sphere, is an example of $PT$
symmetry in the standard sense~\cite{KKM2012Structure}.

Finally, we mention that this method of contractions is quite general.
It applies to all special functions that arise from these systems via separation of variables, not just
polynomials of hypergeometric type, and it extends to higher dimensions,~\cite{KMPost11}.
The special functions arising from the models can be described as the coef\/f\/icients in the expansion of
a~separable eigenbasis for the original quantum system in terms of another separable eigenbasis.
The functions in the Askey scheme are all hypergeometric polynomials that arise as the expansion
coef\/f\/icients relating two separable eigenbases that are {\it both} of hypergeometric type.
Thus, as described in Sections~\ref{Section5} and~\ref{Section6},
there are some contractions which do not f\/it in the Askey
scheme since the physical system fails to have such a~pair of separable eigenbases.
There are also contractions of ${\rm S}9$ to systems that admit 3 independent symmetry operators and are related
to the Askey scheme but are such that the metric becomes singular.
We refer to these systems as ``singular'' and treat them in Section~\ref{Section7}.

The paper is organized as follows.
In Section~\ref{Section2} we give a~brief introduction to superintegrable systems and list the equivalence class
of the physical systems.
In Section~\ref{contractions}, we describe ``natural'' contractions of quadratic algebras.
Section~\ref{Section4} describes the model for ${\rm S}9$ given in terms of the Wilson/Racah polynomials.
Sections~\ref{Section5}, \ref{Section6}, \ref{Section7} and~\ref{Section8} contain the contractions.
Section~\ref{Section9} contains some concluding remarks.
We use the notation of~\cite{AAR} to express all of the orthogonal polynomials in this paper.

\section{Superintegrable systems}\label{Section2}

Now we provide more detail about 2nd order superintegrable systems in 2D.
In local coordinates~$x_i$, the Hamiltonian takes the form $H=\Delta_2+V(\x)$ where
$\Delta_2=\frac{1}{\sqrt{g}}\sum\limits_{ij=1}^2\partial_i(g^{ij}\sqrt{g}\partial_j)$ is the
Laplace--Beltrami operator in these coordinates, $g^{ij}(\x)$ is the contravariant metric tensor and
$g$ is the determinant of the covariant metric tensor.
A 2nd order symmetry operator for this system is a~partial dif\/ferential operator
$L=\frac{1}{\sqrt{g}}\sum\limits_{ij=1}^2\partial_i(L^{ij}(\x)\sqrt{g}\partial_j)+W(\x)$, where
$L^{ij}$ is a~symmetric contravariant tensor, such that $[H,L]\equiv HL-LH=0$.
These operators are formally self-adjoint with respect to the bilinear product
$\langle f_1,f_2\rangle_g=\int f_1(\x)f_2(\x)\sqrt{g(\x)}\,\D x_1\D x_2$ on the manifold~\cite{KMPost10}.
The system is {\it $2$nd order superintegrable} if there are two symmetry operators $L_1$, $L_2$ such that the
set $\{H,L_1,L_2\}$ is algebraically independent, i.e., there is no nontrivial polynomial $P(H,L_1,L_2)$,
symmetric in $L_1$, $L_2$ such that $P\equiv 0$.
Note that if there is only one symmetry operator $L_1$ then the system is 2nd order integrable.
The requirement that two symmetry operators exist is highly restrictive.
It turns out for our treatment of 2nd order 2D quantum superintegrable systems the values of the mass $m$
and Planck's constant $\hbar$ are immaterial, so we have normalized our Hamiltonians as given.

Since every 2D Riemannian space is conformally f\/lat, we can always assume the existence of Cartesian-like
coordinates $x_1, x_2$ such that
\begin{gather*}
H=\frac{1}{\lambda(\x)}(\partial_{11}+\partial_{22})+V(\x),
\qquad
L_k=\frac{1}{\lambda(\x)}\sum_{i,j=1}^2\partial_{i}\big(L^{ij}_{(k)}\lambda\partial_j\big)+W_{(k)}(\x
),\qquad k=1,2.
\end{gather*}
The commutation relations $[H,L_k]=0$, $k=1,2$, put conditions on the potentials $W_{(1)}$, $W_{(2)}$,
enabling us to solve for the partial derivatives $\partial_iW_{(k)}$ in terms of the function $V$ and its
1st derivatives.
The integrability conditions $\partial_1(\partial_2W_{(k)})=\partial_2(\partial_1W_{(k)})$, the
Bertrand--Darboux equations~\cite{KKM20041} lead to the necessary and suf\/f\/icient condition that $V$ must
satisfy a~pair of coupled linear equations of the form
\begin{gather}\label{canonicalequations}
V_{22}-V_{11}=A^{22}V_1+B^{22}V_2,
\qquad
V_{12}=A^{12}V_1+B^{12}V_2,
\end{gather}
for locally analytic functions $A^{ij}(\x)$, $B^{ij}(\x)$.
Here $V_i=\partial_iV$, etc.
We call these the canonical equations.
If the integrability equations for~\eqref{canonicalequations} are satisf\/ied identically then the solution
space for the canonical equations is 4-dimensional and we can always express the general solution in the
form $V(\x)= a_1V_{(1)}(\x)+a_2V_{(2)}(\x)+a_3V_{(3)}(\x)+a_4$ where $a_4$ is a~trivial
additive constant.
In this case we say that the potential is {\it nondegenerate} and refer to it as 3-parameter.
Another possibility is that the solution space is 2-dimensional with general solution $V(\x)=
a_1V_{(1)}(\x)+a_2$. In this case we say that the potential is {\it degenerate} and refer to it as
1-parameter.
Every degenerate potential can be obtained from some nondegenerate potential by restricting the parameters.
It is not {\it just} a~restriction, however, because the structure of the symmetry algebra changes.
A formally skew-adjoint 1st order symmetry may appear and this induces a~new 2nd order symmetry.
The last possibility is that the integrability conditions are satisf\/ied only by a~constant potential.
In that case we refer to the system as {\it free}; the equation $H\Psi=E\Psi$ is just the Laplace--Beltrami
eigenvalue equation.
The case of a~two-parameter potential doesn't occur, i.e., any 2-parameter potential extends to
a~3-parameter potential~\cite{KKMP2009}.

All of these systems have the remarkable property that the symmetry algebras generated by $H$, $L_1$, $L_2$ for
nondegenerate potentials close under commutation.
Def\/ine the 3rd order commutation $R$ by $R=[L_1,L_2]$.
Then the fourth order operators $[R,L_1]$, $[R,L_2]$ are contained in the associative algebra of symmetrized
products of the generators~\cite{KKM20041}:
\begin{gather*}
[L_j,R]=\sum_{0\leq e_1+e_2+e_3\leq2}M^{(j)}_{e_1,e_2,e_3}\big\{L_1^{e_1},L_2^{e_2}\big\}H^{e_3},
\qquad
e_k\geq0,\qquad L_k^0=I,
\end{gather*}
where $\{L_1,L_2\}=L_1L_2+L_2L_1$ is the symmetrizer.
Also the 6th order operator $R^2$ is contained in the algebra of symmetrized products up to 3rd order:
\begin{gather}\label{Casimir1}
R^2-\sum_{0\leq e_1+e_2+e_3\leq3}N_{e_1,e_2,e_3}\big\{L_1^{e_1},L_2^{e_2}\big\}H^{e_3}=0.
\end{gather}
In both equations the constants $M^{(j)}_{e_1, e_2, e_3}$ and $N_{e_1, e_2, e_3}$ are polynomials in the
parameters~$a_1$,~$a_2$, $a_3$ of degree $2-e_1-e_2-e_3$ and $3-e_1-e_2-e_3$, respectively.

For systems with one parameter potentials the situation is dif\/ferent~\cite{KKMP2009}.
There are 4 ge\-nerators: one 1st order $X$ and 3 second order $H$, $L_1$, $L_2$.
The commutators $[X,L_1]$, $[X,L_2]$ are 2nd order and expressed as
\begin{gather}\label{structure2}
[X,L_j]=\sum_{0\leq e_1+e_2+e_3+e_4\leq1}P^{(j)}_{e_1,e_2,e_3,e_4}\big\{L_1^{e_1},L_2^{e_2},X^{2e_3}\big\}H^{e_4},
\qquad
j=1,2,
\end{gather}
where $\{L_1^{e_1},L_2^{e_2}, X^{2e_3+1}\}$ is the symmetrizer of three operators and has 6 terms and
\mbox{$X^0{=}H^0{=}I$}. The commutator $[L_1,L_2]$ is 3rd order, skew adjoint, and expressed as
\begin{gather*}
[L_1,L_2]=\sum_{0\leq e_1+e_2+e_3+e_4\leq1}Q_{e_1,e_2,e_3,e_4}\big\{L_1^{e_1}L_2^{e_2},X^{2e_3+1}\big\}H^{e_4}.
\end{gather*}
Finally, since there are at most 3 algebraically independent generators, there must be a~polynomial
identity satisf\/ied by the 4 generators.
It is of 4th order:
\begin{gather}\label{Casimir2}
G\equiv\sum_{0\leq e_1+e_2+e_3+e_4\leq2}S_{e_1,e_2,e_3,e_4}\big\{L_1^{e_1},L_2^{e_2},X^{2e_3}\big\}H^{e_4}=0.
\end{gather}
The constants $P^{(j)}_{e_1, e_2, e_3, e_4}$, $Q_{e_1, e_2, e_3, e_4}$ and $S_{e_1, e_2, e_3, e_4}$ are
polynomials in the parameter $a_1$ of degrees $1-e_1-e_2-e_3-e_4$, $1-e_1-e_2-e_3-e_4$ and
$2-e_1-e_2-e_3-e_4$, respectively.

All of the possibilities have been classif\/ied.
The classif\/ication is simplif\/ied greatly by use of the St\"ackel transform, an invertible structure
preserving mapping from a~superintegrable system on one manifold to a~superintegrable system on another
manifold~\cite{KKM20041,KKMW, KMPost11}.
Thus, if we know the structure equations for~$H$ then we know the structure equations for any system
St\"ackel equivalent to~$H$.
For our study we can restrict ourselves to a~choice of one representative system in each equivalence class.
There are 13 St\"ackel equivalence classes of systems with nonfree potentials but one is an isolated
Euclidean singleton unrelated to the Askey scheme.
In~\cite{KKM20042}, it is shown that every 2nd order 2D superintegrable system is St\"ackel equivalent to
a~constant curvature system, so we will choose our examples in f\/lat space and on complex 2-spheres.
In~\cite{KKMP} all 21 such systems in f\/lat space are determined up to conjugacy under the complex
Euclidean group and all 9 nonzero constant curvature spaces are determined up to conjugacy under the
complex orthogonal group.
(Some of these systems are St\"ackel equivalent to one another~\cite{Kress2007}.) We will use the notation
given there.
There are thus 6 St\"ackel equivalence classes of nondegenerate potentials and 6 of degenerate potentials.

\subsection{Six nondegenerate superintegrable systems}\label{12systems}

In this section, we f\/ix some notation.
Let $s_1^2+s_2^2+s_3^2=1$ be the embedding of the unit 2-sphere in 3D Euclidean space and $z=x+iy$,
$\overline{z}=x-iy$.
Def\/ine $J_3=s_1\partial_{s_2}-s_2\partial_{s_1}$ to be the generator of rotations about the $s_3$ axis,
with $J_1$, $J_2$ obtained by cyclic permutation.
On the Euclidean plane, we shall also use $J_3$ to denote the generator of rotations about the origin.
In complex coordinates, derivatives are expressed as $\partial=\partial_z$,
$\overline{\partial}=\partial_{\overline z}$.
As in the previous section $R=[L_1, L_2]$.

{\bf 1) Quantum $\boldsymbol{{\rm S}9}$.} This quantum superintegrable system is def\/ined as
\begin{gather*}
H=J_1^2+J_2^2+J_3^2+\frac{a_1}{s_1^2}+\frac{a_2}{s_2^2}+\frac{a_3}{s_3^2},
\\
L_1=J_1^2+\frac{a_3s_2^2}{s_3^2}+\frac{a_2s_3^2}{s_2^2},
\qquad
L_2=J_2^2+\frac{a_1s_3^2}{s_1^2}+\frac{a_3s_1^2}{s_3^2}.
\end{gather*}
The algebra is given by
\begin{gather*}
[L_i,R]=4\{L_i,L_k\}-4\{L_i,L_j\}-(8+16a_j)L_j+(8+16a_k)L_k+8(a_j-a_k),
\\
R^2=\frac83\{L_1,L_2,L_3\}-(16a_1+12)L_1^2-(16a_2+12)L_2^2-(16a_3+12)L_3^2
\\
\phantom{R^2=}{}
+\frac{52}{3}\big(\{L_1,L_2\}+\{L_2,L_3\}+\{L_3,L_1\}\big)+\frac13(16+176a_1)L_1+\frac13(16+176a_2)L_2
\\
\phantom{R^2=}{}
+\frac13(16+176a_3)L_3+\frac{32}{3}(a_1+a_2+a_3)
+48(a_1a_2+a_2a_3+a_3a_1)+64a_1a_2a_3.
\end{gather*}
Here, $\{i,j,k\}$ is a~cyclic permutation of $\{1,2,3\}$ and $L_3$ is given by $L_3=H-L_1-L_2-a_1-a_2-a_3$.

{\bf 2) Quantum E1.} The quantum system is def\/ined by
\begin{gather*}
H=\partial_x^2+\partial_y^2+a_1\big(x^2+y^2\big)+\frac{a_2}{x^2}+\frac{a_3}{y^2},
\\
L_1=\partial_y^2+a_1y^2+\frac{a_3}{y^2},
\qquad
L_2=(x\partial_y-y\partial_x)^2+\left(\frac{a_2y^2}{x^2}+\frac{a_3x^2}{y^2}\right).
\end{gather*}
The algebra relations are
\begin{gather}
\left[L_1,R\right]=8L_1H-8L_1^2+16a_1L_2-8a_1(1+2a_2+2a_3),\nonumber
\\
\left[L_2,R\right]=8\{L_1,L_2\}-4L_2H-16(1+a_2+a_3)L_1+8(1+2a_3)H,\nonumber
\\
R^2=\frac{8}{3}\big(\{L_1,L_2,H\}-\{L_1,L_1,L_2\}\big)-(16a_3+12)H^2\nonumber
\\
\phantom{R^2=}{}
-\left(\frac{176}{3}+16a_2+16a_3\right)L_1^2-16a_1L_2^2+\left(\frac{176}3+32a_3\right)L_1H\nonumber
\\
\phantom{R^2=}{}
+\frac{176a_1}{3}L_2-\frac{16a_1}3\left(12a_2a_3+9a_2+9a_3+2\right).\label{E1Structure}
\end{gather}

{\bf 3) Quantum $\boldsymbol{{\rm E}2}$.} The generators are
\begin{gather*}
H=\partial_x^2+\partial_y^2+a_1\big(4x^2+y^2\big)+a_2x+\frac{a_3}{y^2},
\\
{L}_1=\partial_y^2+a_1y^2+\frac{a_3}{y^2},
\qquad
{L}_2=\frac12\big\{(x\partial_y-y\partial_x),\partial_y\big\}-y^2\left(a_1x-\frac{a_2}{4}\right)+\frac{a_3x}{y^2}.
\end{gather*}
The algebra is def\/ined by
\begin{gather*}
[L_1,R]=2a_2L_2+16a_1L_2,
\qquad
\left[L_2,R\right]=-2L_1^2+4L_2^2-4L_2H+2a_2L_3-a_1(8a_3+6),
\\
{R}^2=4L_2^2+4L_1H+16a_1{L}_2^2-2a_2\{{L}_1,{L}_2\}+(12+16a_3)a_1{L}_1
+32a_1^2L_3-a_2^2\left(a_3+\frac34\right).
\end{gather*}

{\bf 4) Quantum $\boldsymbol{{\rm E}3'}$.} The quantum system is def\/ined by
\begin{gather*}
H=\partial_x^2+\partial_y^2+a_1\big(x^2+y^2\big)+a_2x+a_3y+\frac{a_2^2+a_3^2}{4a_1},
\\
L_1=\partial_y^2+a_1y^2+a_3y+\frac{a_3^2}{4a_1},
\qquad
L_2=2\partial_x\partial_y+\frac{2(a_1x+a_2)(2a_1y+a_3)}{2a_1},
\end{gather*}
with algebra relations
\begin{gather}
\left[L_1,R\right]=-4a_1L_2,\nonumber
\qquad
\left[L_2,R\right]=16a_1L_1+8a_1H,\nonumber
\\
R^2=16a_1L_1H-16a_1L_1^2-4a_1L_2^2-16a_1^2.\label{E3'Structure}
\end{gather}

{\bf 5) Quantum E8.} The quantum system is def\/ined by $(\partial=\partial_z$,
$\overline{\partial}=\partial_{\overline z})$:
\begin{gather*}
H=4\partial\overline{\partial}+a_1z\overline{z}+\frac{a_2z}{\overline{z}^3}+\frac{a_3}{\overline{z}^2},
\\
L_1=-\partial^2-\frac{a_1}{4}\overline{z}^2+\frac{a_2}{2\overline{z}^2},
\qquad
L_2=-\left(z\partial-\overline{z}\overline{\partial}\right)^2+\frac{a_2z^2}{\overline{z}^2}+\frac{a_3z}{\overline{z}}.
\end{gather*}
The algebra relations are
\begin{gather*}
\left[L_1,R\right]=-8L_1^2+2a_1a_2,
\qquad
\left[L_2,R\right]=8\{L_1,L_2\}-16L_1-2a_3H,
\\
R^2=8\big\{L_1^2,L_2\big\}-\frac{176}{3}L_1^2-a_3L_1H+a_2H^2-4a_1a_2L_2-\frac{a_1(3a_3^2-4a_2)}{3}.
\end{gather*}

{\bf 6) Quantum E10.} The quantum system is def\/ined by
\begin{gather*}
H=4\partial\overline{\partial}+a_1\left(z\overline{z}-\frac12\overline{z}
^3\right)+a_2\left(z-\frac32\overline{z}^2\right)+a_3\overline{z},
\\
L_1=-\partial^2-\frac{a_1\overline{z}^2}4-\frac{a_2\overline{z}}2+\frac{a_3}{12},
\\
L_2=\big\{z\partial-\overline{z}\overline{\partial},\partial\big\}
-\overline{\partial}^2-\big(2z+\overline{z}^2\big)
\left(\frac{a_1(2z-3\overline{z}^2)}{16}-\frac{a_2\overline{z}}{2}+\frac{a_3}{4}\right).
\end{gather*}
The algebra relations are
\begin{gather*}
\left[L_1,R\right]=2a_1L_1-\frac{a_2^2}{2}-\frac{a_1a_3}{6},
\qquad
\left[L_2,R\right]=24L_1^2+4a_3L_1-2a_1L_2+a_2H,
\\
R^2=-16L_1^3-\frac{a_1}{4}H^2+2a_1\{L_1,L_2\}-2a_2L_1H-4a_3L_1^2
\\
\phantom{R^2=}{}
-\left(a_2^2+\frac{a_1a_3}{3}\right)L_2-\frac{a_2a_3}{3}H-a_1^2+\frac{a_3^3}{27}.
\end{gather*}

\subsection{Six degenerate superintegrable systems}\label{degeneratesystems}

{\bf 7) Quantum S3 (Higgs oscillator).}\label{S3algebra}
The system is the same as ${\rm S}9$ with $a_1=a_2=0$, $a_3=a$.
The symmetry algebra is generated by
\begin{gather*}
X=J_3,
\qquad
L_1=J_1^2+\frac{a s_2^2}{s_3^2},
\qquad
L_2=\frac12(J_1J_2+J_2J_1)-\frac{a s_1s_2}{s_3^2}.
\end{gather*}
The structure relations for the algebra are given by
\begin{gather}
\left[L_1,X\right]=2L_2,
\qquad
[L_2,X]=-X^2-2L_1+H-a,\nonumber
\\
\left[L_1,L_2\right]=-(L_1X+XL_1)-\left(\frac12+2a\right)X,\nonumber
\\
0=\big\{L_1,X^2\big\}+2L_1^2+2L_2^2-2L_1H+\frac{5+4a}{2}X^2-2aL_1-a.\label{S3Structure}
\end{gather}

{\bf 8) Quantum E14.} The system is def\/ined by
\begin{gather*}
H=\partial_x^2+\partial_y^2+\frac{a}{\overline{z}^2},
\qquad
X=\partial,
\qquad
L_1=\frac{i}{2}\big\{z\partial+\overline{z}\overline{\partial},\partial\big\}+\frac{a}{\overline{z}},
\qquad
L_2=\left(z\partial+\overline{z}\overline{\partial}\right)^2+\frac{az}{\overline{z}},
\end{gather*}
with structure equations
\begin{gather*}
[L_1,L_2]=-\lbrace X,L_2\rbrace-\frac{1}{2}X,
\qquad
[X,L_1]=-X^2,
\qquad
\left[X,L_2\right]=2L_1,
\\
L_1^2+XL_2X-b H-\frac{1}{4}X^2=0.
\end{gather*}

{\bf 9) Quantum E6.} The system is def\/ined by
\begin{gather*}
H=\partial_x^2+\partial_y^2+\frac{a}{x^2},
\qquad
X=\partial_{y},
\\
L_1=(x\partial_y-y\partial_x)^2+\frac{ay^2}{x^2},
\qquad
L_2=\frac12\lbrace x\partial_y-y\partial_x,\partial_{x}
\rbrace-\frac{ay}{x^2},
\end{gather*}
with symmetry algebra
\begin{gather*}
[L_1,L_2]=-\{X,L_1\}-\left(2a+\frac12\right)X,
\qquad
[L_2,X]=H-X^2,
\qquad
\left[L_1,X\right]=2L_2,
\\
L_2^2+\frac14\big\{L_1,X^2\big\}+\frac12XL_1X-L_1H+\left(a+\frac34\right)X^2=0.
\end{gather*}

{\bf 10) Quantum E5.} The system is def\/ined by
\begin{gather*}
H=\partial_x^2+\partial_y^2+ax,
\qquad
X=\partial_y,
\qquad
L_1=\partial_{xy}+\frac1{2}ay,
\qquad
L_2=\frac1{2}\lbrace x\partial_y-y\partial_x,\partial_y\rbrace-\frac1{4}ay^2.
\end{gather*}
The structure equations are
\begin{gather*}
[L_1,L_2]=2X^3-HX,
\qquad
[L_1,X]=-\frac{a}{2},
\qquad
[L_2,X]=L_1,
\\
X^4-HX^2+L_1^2+a L_2=0.
\end{gather*}

{\bf 11) Quantum E4.} The system is def\/ined by
\begin{gather*}
H=\partial_x^2+\partial_y^2+a(x+iy),
\qquad
X=\partial_x+i\partial_y,
\\
L_1=\partial_x^2+a x,
\qquad
L_2=\frac{i}{2}\lbrace x\partial_y-y\partial_x,X\rbrace-\frac{a}{4}(x+iy)^2.
\end{gather*}
The structure equations are
\begin{gather*}
[L_1,X]=a,
\qquad
[L_2,X]=X^2,
\qquad
[L_1,L_2]=X^3+HX-\left\lbrace L_1,X\right\rbrace,
\\
X^4-2\left\lbrace L_1,X^2\right\rbrace+2HX^2+H^2+4a L_2=0.
\end{gather*}

{\bf 12) Quantum E3 (isotropic oscillator).} The system is determined by
\begin{gather*}
H=\partial_x^2+\partial_y^2+a\big(x^2+y^2\big),
\qquad
X=x\partial_y-y\partial_x,
\qquad
L_1=\partial_y^2+a y^2,
\qquad
L_2=\partial_{xy}+a x y.
\end{gather*}
The structure equations are
\begin{gather*}
[L_1,X]=2L_2,
\qquad
[L_2,X]=H-2L_1,
\qquad
[L_1,L_2]=-2a X,
\\
L_1^2+L_2^2-L_1H+aX^2-a=0.
\end{gather*}

\subsection{Two free (1st order) quantum superintegrable systems}

{\bf 1) The 2-sphere.} Here $s_1^2+s_2^2+s_3^2=1$ is the embedding of the unit 2-sphere in
Euclidean space, and the Hamiltonian is $H=J_1^2+J_2^2+J_3^2$, where
$J_3=s_1\partial_{s_2}-s_2\partial_{s_1}$ and $J_2$, $J_3$ are obtained by cyclic permutations of $1$, $2$, $3$.
The basis symmetries are $J_1$, $J_2$, $J_3$.
They generate the Lie algebra so(3) with relations $[J_1,J_2]=-J_3$, $[J_2,J_3]=-J_1$, $[J_3,J_1]=-J_2$ and
Casimir~$H$.

{\bf 2) The Euclidean plane.} 
Here $H= \partial_x^2+\partial_y^2$ with basis symmetries $P_1=\partial_x$, $P_2=\partial_y$ and
$M=x\partial_y-y\partial_x$.
The symmetry Lie algebra is e(2) with relations $[P_1,P_2]=0$, $[P_1,M]=P_2$, $[P_2,M]=-P_1$ and Casimir
$H$.

\vspace{-1mm}

\section{Contractions of superintegrable systems}\label{contractions}

We will give a~detailed treatment of contractions in another publication~\cite{KMSH}, but here we just
describe ``natural'' contractions.
Suppose we have a~nondegenerate superintegrable system with generators $H$, $L_1$, $L_2$ and structure
equations~\eqref{Casimir1}, def\/ining a~quadratic algebra $Q$.
If we make a~change of basis to new generators ${\tilde H}$, ${\tilde L_1}$, ${\tilde L_2}$ and parameters
${\tilde a_1}$, ${\tilde a_2}$, ${\tilde a_3}$ such that
\begin{gather*}
\left(
\begin{matrix}
{\tilde L_1}
\\
{\tilde L_2}
\\
{\tilde H}
\end{matrix}
\right)=\left(
\begin{matrix}
A_{1,1}&A_{1,2}&A_{1,3}
\\
A_{2,1}&A_{2,2}&A_{2,3}
\\
0&0&A_{3,3}
\end{matrix}
\right)\left(
\begin{matrix}
L_1
\\
L_2
\\
H
\end{matrix}
\right)+\left(
\begin{matrix}B_{1,1}&B_{1,2}&B_{1,3}
\\
B_{2,1}&B_{2,2}&B_{2,3}
\\
B_{3,1}&B_{3,2}&B_{3,3}
\end{matrix}
\right)\left(
\begin{matrix}
a_1
\\
a_2
\\
a_3
\end{matrix}
\right),
\\[-1ex]
\left(
\begin{matrix}{\tilde a_1}
\\
{\tilde a_2}
\\
{\tilde a_3}
\end{matrix}
\right)=\left(
\begin{matrix}C_{1,1}&C_{1,2}&C_{1,3}
\\
C_{2,1}&C_{2,2}&C_{2,3}
\\
C_{3,1}&C_{3,2}&C_{3,3}
\end{matrix}
\right)\left(
\begin{matrix}a_1
\\
a_2
\\
a_3
\end{matrix}
\right)
\end{gather*}
for some $3\times 3$ constant matrices $A=(A_{i,j})$, $B$, $C$ such that $\det A \cdot \det C\ne 0$, we will have
the same system with new structure equations of the form~\eqref{Casimir1} for ${\tilde R}=[{\tilde
L_1},{\tilde L_2}]$, $[{\tilde L_j},{\tilde R}]$, ${\tilde R}^2$, but with transformed structure constants.
(Strictly speaking, since the space of potentials is 4-dimensional, we should have a~term $a_4$ in the
above expressions.
However, normally, this term can be absorbed into~$H$.) We choose a~continuous 1-parameter family of basis
transformation matrices $A(\epsilon)$, $B(\epsilon)$, $C(\epsilon)$, $0<\epsilon\le 1$ such that $A(1)=C(1)$ is
the identity matrix, $B(1)=0$ and $\det A(\epsilon)\ne 0$, $\det C(\epsilon)\ne 0$.
Now suppose as $\epsilon\to 0$ the basis change becomes singular  (i.e., the limits of $A$, $B$, $C$ either do
not exist or, if they exist do not satisfy $\det A(0)\det C(0)\ne 0$) but the structure equations involving
$A(\epsilon)$, $B(\epsilon)$, $C(\epsilon)$, go to a~limit, def\/ining a~new quadratic algebra $Q'$.
We call $Q'$ a~{\it contraction} of $Q$ in analogy with Lie algebra contractions~\cite{Wigner}.

For a~degenerate superintegrable system with generators $H$, $X$, $L_1$, $L_2$ and structure
equations~\eqref{structure2}, \eqref{Casimir2}, def\/ining a~quadratic algebra $Q$, a~change of basis to new
generators ${\tilde H}$, ${\tilde X}$, ${\tilde L_1}$, ${\tilde L_2}$ and parameter ${\tilde a}$ such that ${\tilde
a} = Ca$, and
\begin{gather*}
\left(
\begin{matrix}{\tilde L_1}
\\
{\tilde L_2}
\\
{\tilde H}
\\
{\tilde X}
\end{matrix}
\right)=\left(
\begin{matrix}A_{1,1}&A_{1,2}&A_{1,3}&0
\\
A_{2,1}&A_{2,2}&A_{2,3}&0
\\
0&0&A_{3,3}&0
\\
0&0&0&A_{4,4}
\end{matrix}
\right)\left(
\begin{matrix}L_1
\\
L_2
\\
H
\\
X
\end{matrix}
\right)+\left(
\begin{matrix}
B_1
\\
B_2
\\
B_3
\\
0
\end{matrix}
\right)a
\end{gather*}
for some $4\times 4$ matrix $A=(A_{i,j})$ with $\det A\ne 0$, complex 4-vector $B$ and constant $C\ne 0$
yields the same superintegrable system with new structure equations of the
form~\eqref{structure2}, \eqref{Casimir2} for $[{\tilde X},{\tilde L_j}]$, $[{\tilde L_1},{\tilde L_2}]$,
and $\tilde G=0$, but with transformed structure constants.
(Again, strictly speaking, since the space of potentials is 2-dimensional, we should have a~constant term
$c'$ in the above expressions but, normally, this term can be absorbed into~$H$.) Suppose we choose
a~continuous 1-parameter family of basis transformation matrices $A(\epsilon)$, $B(\epsilon)$, $C(\epsilon)$,
$0<\epsilon\le 1$ such that $A(1)$ is the identity matrix, $B(1)=0$, $C(1)=1$, and $\det A(\epsilon)\ne 0$,
$C(\epsilon)\ne 0$.
Now suppose as $\epsilon\to 0$ the basis change becomes singular (i.e., the limits of $A$, $B$, $C$ either do
not exist or, exist but do not satisfy $C(0)\det A(0)\ne 0$), but that the structure equations involving
$A(\epsilon)$, $B(\epsilon)$, $C(\epsilon)$ go to a~f\/inite limit, thus def\/ining a~new quadratic algebra $Q'$.
We call $Q'$ a~{\it contraction} of $Q$.

It has been established that all 2nd order 2D superintegrable systems can be obtained from system ${\rm S}9$ by
limiting processes in the coordinates and/or a~St\"ackel transformation, e.g.\
\cite{KKM2007a, KMWP}.
All systems listed in Subsection~\ref{12systems} are limits of ${\rm S}9$.
It follows that the quadratic algebras generated by each system are contractions of the algebra of ${\rm S}9$.
(However, in general an abstract quadratic algebra may not be associated with a~superintegrable system and
a~contraction of a~quadratic algebra associated with one superintegrable system to a~quadratic algebra
associated with another superintegrable system does not necessarily imply that this is associated with
a~coordinate limit process.)

\vspace{1mm}

\section{Models of superintegrable systems}\label{Section4}

A representation of a~quadratic algebra is a~homomorphism of
the algebra into the associative algebra of linear operators on some vector space.
In this paper a~{\it model} is a~faithful representation in which the vector space is a~space of
polynomials in one complex variable and the action is via dif\/ferential/dif\/ference operators acting on
that space.
We will study classes of irreducible representations realized by these models.
Suppose a~superintegrable system with quadratic algebra $Q$ contracts to a~superintegrable system with
quadratic algebra $Q'$ via a~continuous family of transformations indexed by the parameter $\epsilon$.
If we have a~model of an irreducible representation of $Q$ we can try to ``save'' this representation by
passing through a~continuous family of irreducible representations of $Q(\epsilon)$ in the model to obtain
a~representation of $Q'$ in the limit.
We will show that as a~byproduct of contractions to systems from ${\rm S}9$ for which we save representations in
the limit, we obtain the Askey scheme for hypergeometric orthogonal polynomials.
In all the models to follow, the polynomials we classify are eigenfunctions of formally self-adjoint or
formally skew-adjoint operators.
To present compact results we will not derive the weight functions for the orthogonality; they can be found
in~\cite{KLS}.
They can be determined by requiring that the 2nd order operators $H$, $L_1$, $L_2$ are formally self-adjoint and
the 1st order operator $X$ is formally skew-adjoint.
See~\cite{KMPost11} for some examples of this approach.

\subsection{The S9 model} 

There is no dif\/ferential model for ${\rm S}9$ but a~dif\/ference operator model
yielding structure equations for the Racah and Wilson polynomials~\cite{KMPost1}.
Recall that the Wilson polynomials are def\/ined as
\begin{gather}
w_n\big(t^2\big)\equiv w_n(t^2,a,b,c,d)=(a+b)_n(a+c)_n(a+d)_n\nonumber
\\
\phantom{w_n\big(t^2\big)\equiv}{}
\times
{}_4F_{3}\left(
\begin{matrix}
-n,&a+b+c+d+n-1,&a-t,&a+t\nonumber
\\
a+b,&a+c,&a+d
\end{matrix}
;1\right)
\\
\phantom{w_n\big(t^2\big)}{}
=(a+b)_n(a+c)_n(a+d)_n\Phi^{(a,b,c,d)}_{n}\big(t^2\big),\label{Wilson}
\end{gather}
where $(a)_n$ is the Pochhammer symbol and ${}_4F_3(1)$ is a~hypergeometric function of unit argument.
The polynomial $w_n\big(t^2\big)$ is symmetric in $a$, $b$, $c$, $d$.
For the f\/inite dimensional representations the spectrum of $t^2$ is $\{(a+k)^2,\;k=0,1,\dots,m\}$ and the
orthogonal basis eigenfunctions are Racah polynomials.
In the inf\/inite dimensional case they are Wilson polynomials.
They are eigenfunctions for the dif\/ference operator $\tau^*\tau$ def\/ined via{\samepage
\begin{gather*}
\tau=\frac{1}{2t}\big(E_t^{1/2}-E_t^{-1/2}\big),
\\
\tau^*=\frac{1}{2t}\big[(a+t)(b+t)(c+t)(d+t)E_t^{1/2}-(a-t)(b-t)(c-t)(d-t)E_t^{-1/2}\big],
\end{gather*}
with $E_t^AF(t)=F(t+A)$. }

A f\/inite or inf\/inite dimensional bounded below representation is def\/ined by the following operators
\begin{gather*}
L_1=-4\tau^*\tau-2(\alpha_2+1)(\alpha_3+1)+\frac12,
\qquad
L_2=-4t^2+\alpha_1^2+\alpha_3^2-\frac12,
\qquad
H=E,
\end{gather*}
where $a_i=\frac14-\alpha_i^2$. The energy of the system is
\begin{gather*}
E=-4(m+1)(m+1+\alpha_1+\alpha_2+\alpha_3)
+2(\alpha_1\alpha_2+\alpha_1\alpha_3+\alpha_2\alpha_3)+\alpha_1^2+\alpha_2^2+\alpha_3^2-\frac14,
\end{gather*}
and the constants of the Wilson polynomials are chosen as
\begin{gather*}
a=-\frac12(\alpha_1+\alpha_3+1)-m,
\qquad
d=\alpha_2+m+1+\frac12(\alpha_1+\alpha_3+1),
\\
b=\frac12(\alpha_1+\alpha_3+1),
\qquad
c=\frac12(-\alpha_1+\alpha_3+1).
\end{gather*}
Here $n=0,1,\dots,m$ if $m$ is a~nonnegative integer and $n=0,1,\dots$ otherwise.

Taking a~basis as
\begin{gather*}
f_{n,m}\equiv\Phi^{(a,b,c,d)}_{n}\big(t^2\big),
\end{gather*}
we f\/ind the action of the model on the basis is
\begin{gather*}
L_1f_{n,m}=-\left(4n^2+4n[\alpha_2+\alpha_3+1]+2[\alpha_2+1][\alpha_3+1]-\frac12\right)f_{n,m},
\\
L_2f_{n,m}=K(n+1,n)f_{n+1,m}+K(n-1,n)f_{n-1,m}+\left(K(n,n)+\alpha_1^2+\alpha_3^2-\frac12\right)f_{n,m},
\\
Hf_{m,n}=Ef_{n,m},
\end{gather*}
with
\begin{gather*}
K(n+1,n)={\frac{\left(\alpha_{{3}}+1+\alpha_{{2}}+n\right)\left(m-n\right)\left(m-n+\alpha_{{1}}
\right)\left(1+\alpha_{{2}}+n\right)}{\left(\alpha_{{3}}+1+\alpha_{{2}}+2\,n\right)\left(\alpha_{{3}}
+2+\alpha_{{2}}+2\,n\right)}},
\\
K(n-1,n)={\frac{n\left(\alpha_{{3}}+n\right)\left(\alpha_{{1}}+\alpha_{{3}}+1+\alpha_{{2}}
+m+n\right)\left(1+\alpha_{{3}}+\alpha_{{2}}+m+n\right)}{\left(\alpha_{{3}}+1+\alpha_{{2}}
+2\,n\right)\left(\alpha_{{3}}+\alpha_{{2}}+2\,n\right)}},
\\
K(n,n)=\left(\frac12(\alpha_1+\alpha_3+1)-m\right)^2-K(n+1,n)-K(n-1,n).
\end{gather*}

\section{Nondegenerate to nondegenerate limits}\label{Section5}

\subsection[Contractions S9 $\to$ E1]{Contractions S9 $\boldsymbol{\to}$ E1}

There are at least two ways to take this contraction; it is possible
to contract the sphere about the point $(0,1,0)$ which gives the contraction of representation in terms of
Wilson polynomials to continuous dual Hahn polynomials.
Contracting about the point $(1,0,0)$ leads to continuous Hahn polynomials or Jacobi polynomials.
The continuous dual Hahn and continuous Hahn polynomials correspond to the same superintegrable system but
they are eigenfunctions of dif\/ferent generators.
For the f\/inite dimensional restrictions ($m$ a~positive integer) we have the restrictions of Racah
polynomials to dual Hahn and Hahn, respectively.

We would like to mention that the algebra associated with the Hartmann potential (a speciali\-za\-tion of E1)
has already been associated with the Hahn algebra and the overlap coef\/f\/icients of separable solutions
have been expressed in terms of Hahn polynomials~\cite{GZLU}.
As this section shows, the algebra can be directly obtained by the canonical Lie algebra contraction from
so(3) to e(2).

{\bf 1) Wilson $\boldsymbol{\to}$ Continuous dual Hahn.} For the f\/irst limit, in the quantum system,
we contract about the point $(0,1,0)$ so that the points of our two dimensional space lie in the plane
$(x,1, y)$. We set $s_1=\sqrt{\epsilon}x$, $s_2=\sqrt{1-s_1^2-s_3^2}\approx 1-\frac{\epsilon}{2}(x^2+y^2)$,
$s_3=\sqrt{\epsilon }y$, for small $\epsilon$.
The coupling constants are transformed as
\begin{gather*}
\left(
\begin{matrix}{\tilde a_1}
\\
{\tilde a_2}
\\
{\tilde a_3}
\end{matrix}
\right)=\left(
\begin{matrix}0&\epsilon^2&0
\\
1&0&0
\\
0&0&1
\end{matrix}
\right)\left(
\begin{matrix}a_1
\\
a_2
\\
a_3
\end{matrix}
\right),
\end{gather*}
and we get E1 as $\epsilon\to 0$.
This gives the quadratic algebra contraction def\/ined by the contraction of the operators
\begin{gather*}
\left(
\begin{matrix}{\tilde L_1}
\\
{\tilde L_2}
\\
{\tilde H}
\end{matrix}
\right)=\left(
\begin{matrix}
\epsilon&0&0
\\
0&1&0
\\
0&0&\epsilon
\end{matrix}
\right)\left(
\begin{matrix}L_1
\\
L_2
\\
H
\end{matrix}
\right)+\left(
\begin{matrix}
0&0&0
\\
0&0&0
\\
0&-\epsilon&0
\end{matrix}
\right)\left(
\begin{matrix}a_1
\\
a_2
\\
a_3
\end{matrix}
\right).
\end{gather*}
As in ${\rm S}9$, it is advantageous in the model to express the 3 coupling constants as quadratic functions of
other parameters, so that
\begin{gather}\label{betas}
\left(
\begin{matrix}{\tilde a_1}
\\
{\tilde a_2}
\\
{\tilde a_3}
\end{matrix}
\right)=\left(
\begin{matrix}-\beta_1^2
\\
\frac14-\beta_2^2
\vspace{1mm}\\
\frac14-\beta_3^2
\end{matrix}
\right)=\left(
\begin{matrix}\frac{\epsilon^2}4-\epsilon^2\alpha_2^2
\vspace{1mm}\\
\frac14-\alpha_1^2
\vspace{1mm}\\
\frac14-\alpha_3^2
\end{matrix}
\right)
\end{gather}
with $\alpha_2\rightarrow \infty$ to save the representation.

In the contraction limit the operators tend to
\begin{gather}
L_1'=\lim\limits_{\epsilon\rightarrow0}\widetilde{L}_1=-4\tau'^*\tau'-2\beta_1(\beta_3+1),
\qquad
L_2'=\lim\limits_{\epsilon\rightarrow0}\widetilde{L}_2=-4t^2+\beta_2^2+\beta_3^2-\frac12,
\nonumber
\\
H'=\lim\limits_{\epsilon\rightarrow0}\widetilde{H}=E'.\label{E1dualHahn}
\end{gather}
The energy of the system is now
\begin{gather*}
E'=-2\,\beta_{{1}}\left(2\,m+2+\beta_{{2}}+\beta_{{3}}\right).
\end{gather*}
The eigenfunction of $L_1$, the Wilson polynomials, transform in the contraction limit to the
eigenfunctions of $L_1'$, the dual Hahn polynomials $S_n$,
\begin{gather*}
S_n\big(-t^2,a',b',c'\big)=(a'+b')_n(a'+c')_n{}_3F_2\left(
\begin{matrix}
-n,&a'+t,&a'-t
\\
a'+b',&a'+c'&
\end{matrix}
;1\right)
\end{gather*}
where the constants of the dual Hahn polynomials are
\begin{gather*}
a'=-\frac12(\beta_2+\beta_3+1)-m,
\qquad
b'=\frac12(\beta_2+\beta_3+1),
\qquad
c'=\frac12(-\beta_2+\beta_3+1).
\end{gather*}
Again, $n=0,1,\dots,m$ if $m$ is a~nonnegative integer and $n=0,1,\dots$ otherwise.
The opera\-tors~$\tau'^*$ and $\tau'$ are given by
\begin{gather*}
\tau'=\tau=\frac{1}{2t}\big(E_t^{1/2}-E_t^{-1/2}\big),
\\
\tau'^*=\frac{\beta_1}{2t}\big[(a'+t)(b'+t)(c'+t)E_t^{1/2}-(a'-t)(b'-t)(c'-t)E_t^{-1/2}\big].
\end{gather*}

Taking a~basis as
\begin{gather*}
f'_{n,m}\equiv\frac{S_{n}(-t^2,a',b',c')}{(a'+b')_n(a'+c')_n},
\end{gather*}
we f\/ind that the action of the model is
\begin{gather*}
L_1'f'_{n,m}=-2\beta_1\left(2n+\beta_3+1\right)f'_{n,m},
\\
L_2'f'_{n,m}=K'(n+1,n)f'_{n+1,m}+K'(n-1,n)f'_{n-1,m}
+\left(K'(n,n)+\beta_2^2+\beta_3^2-\frac12\right)f_{n,m},
\\
H'f_{m,n}=E'f_{n,m},
\end{gather*}
with
\begin{gather}
K'(n+1,n)=\left(m-n\right)\left(m-n+\beta_2\right),
\qquad
K'(n-1,n)=n(n+\beta_3),\nonumber
\\
K'(n,n)=\left(\frac12(\beta_2+\beta_3+1)-m\right)^2-K'(n+1,n)-K'(n-1,n).\label{E1dualHahnaction}
\end{gather}

{\bf 2) Wilson $\boldsymbol{\to}$ Continuous Hahn.} For the next limit, we contract about the point
$(1,0,0)$ so that the points of our two dimensional space lie in the plane $(1,x,y)$. We set
$s_1=\sqrt{1-s_1^2-s_3^2}\approx 1-\frac{\epsilon}{2}(x^2+y^2)$, $s_2=\sqrt{\epsilon} x$,
$s_3=\sqrt{\epsilon }y$, for small $\epsilon$.
The coupling constants are transformed as
\begin{gather*}
\left(
\begin{matrix}
{\tilde a_1}
\\
{\tilde a_2}
\\
{\tilde a_3}
\end{matrix}
\right)=\left(
\begin{matrix}
\epsilon^2&0&0
\\
0&1&0
\\
0&0&1
\end{matrix}
\right)\left(
\begin{matrix}a_1
\\
a_2
\\
a_3
\end{matrix}
\right),
\end{gather*}
and we get E1 as $\epsilon\to 0$.
This gives the quadratic algebra contraction
\begin{gather}\label{SecondS9-E1contraction}
\left(
\begin{matrix}{\tilde L_1}
\\
{\tilde L_2}
\\
{\tilde H}
\end{matrix}
\right)=\left(
\begin{matrix}
0&\epsilon&0
\\
1&0&0
\\
0&0&\epsilon
\end{matrix}
\right)\left(
\begin{matrix}L_1
\\
L_2
\\
H
\end{matrix}
\right)+\left(
\begin{matrix}
0&0&0
\\
0&0&0
\\
-\epsilon&0&0
\end{matrix}
\right)\left(
\begin{matrix}a_1
\\
a_2
\\
a_3
\end{matrix}
\right).
\end{gather}
In terms of the constants of~\eqref{betas} the transformation gives
\begin{gather*}
\left(
\begin{matrix}{\tilde a_1}
\\
{\tilde a_2}
\\
{\tilde a_3}
\end{matrix}
\right)=\left(
\begin{matrix}-\beta_1^2
\\
\frac14-\beta_2^2
\vspace{1mm}\\
\frac14-\beta_3^2
\end{matrix}
\right)=\left(
\begin{matrix}\frac{\epsilon^2}4-\epsilon^2\alpha_1^2
\vspace{1mm}\\
\frac14-\alpha_2^2
\vspace{1mm}\\
\frac14-\alpha_3^2
\end{matrix}
\right),
\end{gather*}
with $\alpha_1\rightarrow \infty$.

Saving a~representation: We set
\begin{gather*}
t=-x+{\frac{\beta_1}{2\epsilon}}+m+\frac12(\beta_{{3}}+1).
\end{gather*}
In the contraction limit, the operators are def\/ined as $L_i'=\lim\limits_{\epsilon\rightarrow 0}\widetilde{L}_i$
with
\begin{gather}
L_1'=2\beta_{{1}}\left(2\,x-2m-\beta_{{3}}-1\right),
\nonumber
\\
L_2'=-4\left(B(x)E_x+C(x)E_x^{-1}-B(x)-C(x)\right)-2(\beta_2+1)(\beta_3+1)+\frac12,
\nonumber
\\
H'=-2\beta_1(2m+2+\beta_2+\beta_3),\label{E1Hahn}
\end{gather}
where $B(x)=(x-m)(x+\beta_2+1), \, C(x)=x(x-m-1-\beta_3)$. The operators $L_1'$, $L_2'$ and $H'$ satisfy the
algebra relations in~\eqref{E1Structure}.
The eigenfunction of $L_1$, the Wilson polynomials, transform in the contraction limit to the
eigenfunctions of $L_2'$, which are the Hahn polynomials, ${f'}_{m,n}=Q_n$,
\begin{gather*}
Q_n(x;\beta_2,\beta_3,m)={}_3F_2\left(
\begin{matrix}
-n,&\beta_2+\beta_3+n+1,&-x
\\
-m,&\beta_2+1&
\end{matrix}
;1\right).
\end{gather*}
The action of the operators on this basis is given by
\begin{gather*}
L_1'f'_{n,m}=K'(n+1,n)f'_{n+1,m}+K'(n,n)f'_{n,m}+K'(n-1,n)f'_{n-1,n},
\\
L'_2f'_{n,m}=-\left(4n^2+4n[\beta_2+\beta_3+1]+2[\beta_2+1][\beta_3+1]-\frac12\right)f'_{n,m},
\\
K'(n+1,n)=-4\beta_1\frac{(m-n)(n+\beta_2+\beta_3+1)(n+\beta_2+1)}
{(2n+\beta_2+\beta_3+1)(2n+\beta_2+\beta_3+2)},
\\[0.75mm]
K'(n-1,n)=-4\beta_1\frac{n(n+\beta_3)(m+n+\beta_2+\beta_3+1)}{(2n+\beta_2+\beta_3+1)(2n+\beta_2+\beta_3)},
\\[0.75mm]
K'(n,n)=-2\beta_1(2m+\beta_3+1)-K'(n+1,n)-K(n-1,n).
\end{gather*}

If $\beta_2={\bar \beta_3}$ there is still a~real 3-term recurrence relation.
In the original quantum system the potential is PT-symmetric even though complex, so the energy spectrum is
real.
In this case one studies the original system and its dual and obtains biorthogonality, rather than an
orthonormal basis.

{\bf 3) Wilson $\boldsymbol{\to}$ Jacobi.} The previous contraction is undef\/ined when
$\alpha_1=\beta_1=0$.
However, we can save this representation by setting
\begin{gather*}
m=\frac{\sqrt{-E'}}{2\sqrt{\epsilon}}-1+\frac{\beta_2+\beta_3}{2},
\qquad
t=\frac{\sqrt{-E'}}{2\sqrt{\epsilon}}\sqrt{\frac{1+x}{2}},
\end{gather*}
for $E'$ a~constant and letting $m\to\infty$.
Then, the contraction~\eqref{SecondS9-E1contraction} gives a~contraction of the model for ${\rm S}9$ to
a~dif\/ferential operator model for E1 with $\beta_1=0$:
\begin{gather}
H'=\!E',
\qquad
L_1'=\!\frac{E'}{2}(x+1),\nonumber
\\
L_2'=\!4\big(1-x^2\big)\partial_x^2
+4\big[\beta_3-\beta_2-(\beta_2+\beta_3+2)x\big]\partial_x-2(\beta_2+1)(\beta_3+1)+\frac12.\label{E1Model3}
\end{gather}
The eigenfunctions for $L_1$, the Wilson polynomials, tend in the limit to eigenfunction of $L_2'$ which
are the Jacobi polynomials:
\begin{gather}\label{Jacobia}
P_n^{\beta_2,\beta_3}(x)=\frac{(\beta_2+1)_n}{n!}{}_2F_1\left(
\begin{matrix}
-n,&\beta_2+\beta_3+n+1
\\
\beta_2+1&
\end{matrix}
;\frac{x-1}{2}\right).
\end{gather}
Taking a~basis as $f_{n}=\frac{n!}{(\beta_2+1)_n}P_n^{\beta_2,\;\beta_3}(x)$, we f\/ind that the action of
the operators is
\begin{gather*}
L_1'f'_{n}=K'(n+1,n)f'_{n+1}+K'(n-1,n)f_{n-1}+K'(n,n)f_{n},
\\
L_2'f'_n=-4n(n+\beta_2+\beta_3+1)-2(\beta_2+1)(\beta_3+1)+\frac12,
\end{gather*}
with
\begin{gather*}
K'(n+1,n)=\frac{E'(\beta_2+\beta_3+n+1)(\beta_2+n+1)}{(\beta_2+\beta_3+2n+1)(\beta_2+\beta_3+2n+2)},
\\[1mm]
K'(n-1,n)=\frac{E'n(n+\beta_3)}{(\beta_2+\beta_3+2n)(\beta_2+\beta_3+2n+1)},
\\[1mm]
K'(n,n)=E'-K'(n+1,n)-K'(n-1,n).
\end{gather*}

In this case the basis functions above, suitably renormalized, are pseudo Jacobi polyno\-mials~\cite{Askey}.
If $\beta_2={\bar \beta_3}$ there is a~real 3-term recurrence relation and the potential is PT-symmetric so
$E$ is real.
Then one studies the original system and its dual and obtains biorthogonality of the basis functions.

\subsection[Contractions E1 $\to$ E8]{Contractions E1 $\boldsymbol{\to}$ E8}

In the contraction limit from E1 to E8, the Jacobi polynomials are obtained.
This is in agreement with the fact that the E8 structure algebra coincides with the quadratic Jacobi
algebra QJ(3) def\/ined in~\cite{Zhedanov1992a}.

{\bf 1) Hahn $\boldsymbol{\to}$ Jacobi.} Here, we express 2D Euclidean space in complex
variables and let $z \to \infty, \, \overline{z} \to 0$ as
\begin{gather*}
x=\frac12\big(\sqrt{\epsilon}z+\overline{z}\epsilon^{-\frac12}\big),
\qquad
y=\frac{-i}{2}\big(\sqrt{\epsilon}z-\overline{z}\epsilon^{-\frac12}\big).
\end{gather*}
The coupling constants are transformed as
\begin{gather*}
\left(
\begin{matrix}{\tilde a_1}
\\
{\tilde a_2}
\\
{\tilde a_3}
\end{matrix}
\right)=\left(
\begin{matrix}
1&0&0
\\
0&-8\epsilon^2&-8\epsilon^2
\\
0&4\epsilon&-4\epsilon
\end{matrix}
\right)\left(
\begin{matrix}a_1
\\
a_2
\\
a_3
\end{matrix}
\right),
\end{gather*}
and the system E8 is obtained by the following singular limit:
\begin{gather}\label{E1-E8contraction}
\left(
\begin{matrix}{\tilde L_1}
\\
{\tilde L_2}
\\
{\tilde H}
\end{matrix}
\right)=\left(
\begin{matrix}
\epsilon&0&0
\\
0&1&0
\\
0&0&1
\end{matrix}
\right)\left(
\begin{matrix}L_1
\\
L_2
\\
H
\end{matrix}
\right)+\left(
\begin{matrix}
0&0&0
\\
0&1&1
\\
0&0&0
\end{matrix}
\right)\left(
\begin{matrix}a_1
\\
a_2
\\
a_3
\end{matrix}
\right).
\end{gather}

Consider the second E1 model above, based on the Hahn polynomials.
For simplicity of the model, we introduce new parameters $\gamma_i$ as in
\begin{gather}\label{gammas}
\left(
\begin{matrix}{\tilde a_1}
\\
{\tilde a_2}
\\
{\tilde a_3}
\end{matrix}
\right)=\left(
\begin{matrix}-\gamma_1^2
\\
16\gamma_3^2+\mathcal{O}(\epsilon)
\\
8\gamma_3(\gamma_2-2)+\mathcal{O}(\epsilon)
\\
\end{matrix}
\right)=\left(
\begin{matrix}-\beta_1^2
\\
-4\epsilon^2(1-2\beta_2^2-2\beta_3^2)
\\
-4\epsilon\big(\beta_2^2-\beta_3^2\big)
\end{matrix}
\right).
\end{gather}
We save the representation via~\eqref{gammas} and the following change of variable
\begin{gather*}
x=\frac{\gamma_3}{\epsilon}\left(\frac{1-t}{2}\right).
\end{gather*}
In the contraction limit $L_i'=\lim\limits_{\epsilon\rightarrow 0}\tilde{L}_i$~\eqref{E1-E8contraction}, the model
becomes
\begin{gather*}
L_1'=-2\gamma_1\gamma_3t,
\qquad
L_2'=4\big(1-t^2\big)\partial_t^2+4(2m+\gamma_2-\gamma_2t)\partial_t-(\gamma_2-1)^2,
\\
H'=-2\gamma_1(2m+\gamma_2).
\end{gather*}
The eigenfunctions for $L_2$~\eqref{E1Hahn}, Hahn polynomials, tend in the limit to eigenfunctions of
$L_2'$, Jacobi polynomials
\begin{gather*}
P_{n}^{-m-1,\gamma_2+m-1}(t)=\frac{(-m)_n}{n!}\, {}_2F_1\left(
\begin{matrix}
-n&n+\gamma_2+1
\\
&-m
\end{matrix}
;\frac{1-t}{2}\right),
\end{gather*}
with the normalization $\lim\limits_{\epsilon\rightarrow 0}f_{m,n}=f'_{m,n}=\frac{n!}{ (-m)_n}P_{n}^{-m-1,
\gamma_2+m-1}(t)$. The action of the operators on this basis becomes
\begin{gather*}
L_1'f'_{n}=K'(n+1,n)f'_{n+1}+K'(n-1,n)f_{n-1}+K'(n,n)f_{n},
\\
L_2'f'_n=\left[-4n(n+\gamma_2-1)-(\gamma_2-1)^2\right]f'_n,
\end{gather*}
with
\begin{gather*}
K'(n+1,n)=-\frac{4\gamma_1\gamma_3(m-n)(n+\gamma_2-1)}{(2n+\gamma_2-1)(2n+\gamma_2)},
\\
K'(n-1,n)=\frac{4\gamma_1\gamma_3n(n+m+\gamma_2-1)}{(2n+\gamma_2-1)(2n+\gamma_2-2)},
\\
K(n,n)=-2\gamma_1\gamma_3-K(n+1,n)-K(n-1,n).
\end{gather*}
Note that this model gives a~f\/inite dimensional representation in the case that $m$ is a~positive integer.
This is in contrast to the previous model based on Jacobi polynomials~\eqref{Jacobia} which gives only
inf\/inite dimensional representations and in agreement with the fact that the classical physical system E1
with $a_1=0$ has only unbounded trajectories.

{\bf 2) Jacobi $\boldsymbol{\to}$ Generalized Bessel polynomials.} The same
contraction~\eqref{E1-E8contraction} acting on the model for E1 with $a_1=0$~\eqref{E1Model3}, gives
a~model based on the generalized Bessel polynomials~\cite{KF1949}:
\begin{gather*}
L_1'=-\gamma_3E't,
\qquad
L_2'=-4t^2\partial_t^2-4(1+\gamma_2t)\partial_t-(\gamma_2-1)^2,
\qquad
H'=E',
\end{gather*}
where we have used the change of variable $x=-2\gamma_3t/\epsilon$.

\subsection[Contraction ${\rm E}1\to{\rm E}3'$]{Contraction $\boldsymbol{{\rm E}1\to {\rm E}3'}$}

{\bf 1) Dual Hahn $\boldsymbol{\to}$ Meixner, Krawtchouk, and Meixner--Pollaczek.}
For this contraction, we make a~contraction which is not ``natural'' in the sense of
Section~\ref{contractions}.
Beginning with the quantum E1 system, the change of variables
\begin{gather*}
x\rightarrow x+\sqrt{\frac{2c_2}{\epsilon\sqrt{-a'_1}}}+\frac{a_2'}{2a_1'},
\qquad
y\rightarrow y+\sqrt{\frac{2c_1}{\epsilon\sqrt{-a'_1}}}+\frac{a_3'}{2a_1'}
\end{gather*}
has a~f\/inite limit for the following change of parameters
\begin{gather*}
a_1=\frac14a_1',
\qquad
a_2=-\frac{c_1^2}{\epsilon^2},
\qquad
a_3=-\frac{c_2^2}{\epsilon^2},
\end{gather*}
and operators
\begin{gather*}
H'=\lim\limits_{\epsilon\rightarrow0}H+\frac{(c_1+c_2)\sqrt{-a_1'}}{\epsilon}
=\partial_x^2+\partial_y^2+a_1'\big(x^2+y^2\big)+a_2'x++a_3'y+\frac{(a_2')^2+(a_3')^2}{4a_1'},
\\
L_1'=\lim\limits_{\epsilon\rightarrow0}L_1+\frac{c_1\sqrt{-a_1'}}{\epsilon}
=\partial_y^2+a_1'y^2+a_3'y+\frac{(a_3')^2}{4a_1'},
\\
L_2'=-\frac{\epsilon}{4}\sqrt{\frac{-a_1'}{c_1c_2}}L_2-\frac{\epsilon}{2}\sqrt{-a_1'c_1c_2}
=\partial_x\partial_y+a_1'xy+\frac{a_2'}{2}x+\frac{a_3'}{2}y+\frac{a_2'a_3'}{4a_1'}
\\
\phantom{L_2'=}{}
+\frac{c_2}{2\sqrt{c_1c_2}}H'+\frac{c_1-c_2}{2\sqrt{c_1c_2}}L_1'.
\end{gather*}
It's clear that these operators generate the algebra E$3'$~\eqref{E3'Structure}.

In terms of the constants used in the f\/irst model for E1~\eqref{E1dualHahn}, the $\beta_i$, become
\begin{gather*}
\beta_1=\sqrt{-a_1}=\frac12\omega,
\qquad
\beta_2=\frac{c_1}{\epsilon}+\mathcal{O}(\epsilon),
\qquad
\beta_3=\frac{c_2}{\epsilon}+\mathcal{O}(\epsilon).
\end{gather*}
Here, we have introduced the new constant $\omega$.
The model~\eqref{E1dualHahn} has a~f\/inite limit under the change of variable
$t=x-m-1/2+\epsilon^{-1}(c_1+c_2)/2$. The following operators thus form a~model for the E$3'$ algebra:
\begin{gather*}
H'=-2\omega(m+1),
\qquad
L_1'=\frac{2\omega c_2}{c_1+c_2}\left[B(x)E_x+C(x)E_x^{-1}-(B(x)+C(x))\right]-\omega,
\\
B(x)=\left(-\frac{c_1}{c_2}\right)(x-m),
\qquad
C(x)=x,
\qquad
L_2'=\frac{\omega(c_1+c_2)}{\sqrt{c_1c_2}}\left(2x-2m+1\right).
\end{gather*}
The eigenfunctions of $L_1'$ are given by Meixner polynomials
\begin{gather}\label{MeixnerPollaczek}
f'_{n,m}={}_2F_1\left(
\begin{matrix}
-n,&-x
\\
-m,&
\end{matrix}
;1-\frac{1}{c}\right),
\end{gather}
which have been obtained as limits of Hahn polynomials.
Here, $c=-c_1/c_2$.
In the case where~$m$ is a~positive integer the Meixner polynomials reduce to Krawtchouk polynomials.

The action of the model on this basis is given by
\begin{gather*}
L_1'f'_{n,m}=-\omega(2n+1),
\\
L_2'f'_{n,m}=K'(n+1,n)f'_{n+1,m}+K'(n-1,n)f'_{n-1,m}+K'(n,n)f'_{n,m},
\end{gather*}
with
\begin{gather*}
K'(n+1,n)=\frac{2\omega(m-n)c_1}{\sqrt{c_1c_2}},
\qquad
K'(n-1,n)=\frac{-2\omega n c_2}{\sqrt{c_1c_2}},
\\
K'(n,n)=\frac{\omega(c_1+c_2)(2m+1)}{\sqrt{c_1c_2}}-K'(n+1,n)-K'(n-1,n),
\end{gather*}
which agrees with the limit of the action of the E1 model on the dual Hahn basis~\eqref{E1dualHahnaction}.

Recall that in the model for the system E1, the dual Hahn polynomials had a~real 3-term recurrence relation
when the system itself was $PT$-symmetric.
If we retain this restriction in the limit, the constant $c$ is required to have modulus $1$, $c=e^{2i\phi}$,
and $a_2'=\overline{a_3'}$ in the physical system.
In this case, the Meixner--Pollaczek polynomials are obtained as a~limit of the dual Hahn
\begin{gather*}
P_n^\lambda(x;\phi)=\frac{(-m)_n}{n!}e^{in\phi}\, {}_2F_1\left(
\begin{matrix}
-n,&-\frac{m}{2}+ix
\\
-m,&
\end{matrix}
;1-e^{-2i\phi}\right).
\end{gather*}
Here, we have made a~change of variables $x\rightarrow ix+m/2$.

The related E3$'$ quantum system has special properties.
We choose it as
\begin{gather*}
H=\partial_{xx}+\partial_{yy}-\omega^2\big(x^2+y^2\big)+a_2'x+\overline{a_2}'y-\frac{a_2'^2+\overline{a_2'}^2}
{4\omega^2},
\end{gather*}
where $\omega>0$.
This system admits $PT$-symmetry; the potential $V$ is complex but the bound-state eigenvalues are real:
\begin{gather*}
E_m=-2\omega(n_1+n_2+1),
\qquad
n_1+n_2=m=0,1,2,\dots.
\end{gather*}
Here,~$H$ is not self-adjoint but its basis vectors and the basis vectors of its adjoint $H^*$ are
biorthogonal.

{\bf 2) Hahn $\boldsymbol{\to}$ Meixner, Krawtchouk, and Meixner--Pollaczek.}
We use the same limit as immediately above, but apply it to the second E1
model~\eqref{E1Hahn} to obtain
\begin{gather*}
H'=-2\omega(m+1),
\qquad
L_1'=-\omega(2x-2m-1),
\qquad
L_2'=\frac{2\omega}{\sqrt{-c}}\left[(x-m)T^{1}_x+2T^{-1}_x\right].
\end{gather*}
The operator which is diagonalized by the Meixner polynomials~\eqref{MeixnerPollaczek} is
$c\sqrt{-c}L_2'+(1+c)L_1-cH'$. The above discussion of the $PT$-symmetric limit also applies for this model.

{\bf 3) Krawtchouk $\boldsymbol{\to}$ Charlier.}
Beginning with the Krawtchouk basis model~\eqref{MeixnerPollaczek} we consider the $m\rightarrow\infty$ limit.
This is a~dif\/ferent type of contraction than considered above because the $m+1$ dimensional eigenspace is
changing with each increment of $m$.
We can save the representation by taking $c_2=1/m$, $c_1=-(1/a+1/m)$.
The basis functions become
\begin{gather*}
f_n(x)=C_n(x;a)=\lim\limits_{m\to\infty}{}_2F_1\left(
\begin{matrix}
-n,&-x
\\
m
\end{matrix}
;-\frac{m}{a}\right)={}_2F_0\left(
\begin{matrix}
-n,&-x
\\
-
\end{matrix}
;-\frac{1}{a}\right),
\end{gather*}
the Charlier polynomials.
The dif\/ference operator determining these polynomials and the three term recurrence relation are obtained
in the limit:
\begin{gather*}
{\hat L_2}'f_{n}=(L_2'+4\omega m)f_{n}=4x\omega f_{n}=-4K(n+1,n)f_{n+1}-4K(n,n)f_{n}-4K(n-1,n)f_{n-1},
\\[1.75mm]
K(n+1,n)=-\omega a,
\qquad
K(n-1,n)=\omega n,
\qquad
K(n,n)=\omega(a-n),
\\
L_3f_{n}=-\frac{4n}{a}f_{n}=-4\left(B(x)T_x+C(x)T_x^{-1}-[B(x)+C(x)]\right)f_{n},
\\
B(x)=-1,
\qquad
C(x)=\frac{x}{a},
\qquad
{\hat H'}=H'+4\omega m=0.
\end{gather*}
As we go to the contraction limit the model is restricted to the eigenspace of $H'$ with eigenvalue
$-4\omega m$, i.e., on the eigenspace of ${\hat H'}$ with eigenvalue 0.
In the quantum system, the Hamiltonian $H'+4\omega m$ and the symmetry $L_2'+4\omega m$ blow up with $m$,
so don't give a~f\/inite limit.
However, the quadratic algebra converges to itself in this contraction of E3$'$.
The model is giving us asymptotic information in $m$ about the relation between $L_2'$ and $L_1'$
eigenbases of $H'$ on the $-4\omega m$ eigenspace.

\subsection[Contraction E1 $\to$ E2]{Contraction E1 $\boldsymbol{\to}$ E2}

In quantum E1 we let $x\rightarrow x+\sqrt{\frac{c}{\epsilon
\sqrt{-a'_1}}} +\frac{a_2'}{8a_1'}$, $y\to y$, and go to the limit.
This induces a~algebra contraction to ${\rm E}2$.
Setting $a_2=-c^2/\epsilon^2$, $H'=\lim\limits_{\epsilon\rightarrow 0}H+2\sqrt{-a_1'}c/\epsilon$,
${L}_1'={L}_1$, and ${L}_2'=\frac12\left(\frac{c}{\epsilon
\sqrt{-a'_1}}\right)^{-\frac12}L_2-\frac12\left(\left(\frac{c}{\epsilon
\sqrt{-a'_1}}\right)^{\frac12}+\frac{a_2'}{4a_1'}\right)L_1$, gives a~contraction to ${\rm E}2$.
We can't save the representation.
Using other models we can show that this contraction yields information about limits of non-Gaussian
hypergeometric functions, not related to the Askey scheme.

\subsection[Contraction E8 $\to$ E10]{Contraction E8 $\boldsymbol{\to}$ E10}

In the physical model we translate to inf\/inity: $z=z'-1/\epsilon^2$,
${\bar z}={\bar z}'+1/\epsilon$ and cancel the singularities that occur.
The coupling constants transform as
\begin{gather*}
\left(
\begin{matrix}{\tilde a_1}
\\
{\tilde a_2}
\\
{\tilde a_3}
\end{matrix}
\right)=\left(
\begin{matrix}
\frac{-4a_1}{3}&0&-\frac{8a_3\epsilon^5}{3}
\\
\frac{4a_1}{3\epsilon}&0&\frac{2a_3\epsilon^4}{3}
\\
\frac{-a_2}{\epsilon^2}&3\epsilon^2a_2&-2\epsilon^3a_3
\end{matrix}
\right)\left(
\begin{matrix}a_1
\\
a_2
\\
a_3
\end{matrix}
\right),
\end{gather*}
and the system E10 is obtained by the following singular limit:
\begin{gather*}
\left(
\begin{matrix}{\tilde L_1}
\\
{\tilde L_2}
\\
{\tilde H}
\end{matrix}
\right)=\left(
\begin{matrix}
1&0&0
\\
-\frac{1}{\epsilon^2}&\epsilon^2&\frac{1}{2\epsilon}
\\
0&0&1
\end{matrix}
\right)\left(
\begin{matrix}L_1
\\
L_2
\\
H
\end{matrix}
\right)+\left(
\begin{matrix}
\frac{1}{6\epsilon^2}&0&-\frac{\epsilon^3}{6}
\vspace{1mm}\\
\frac{1}{4\epsilon^4}&-\frac14&\frac{\epsilon}{2}
\vspace{1mm}\\
\frac{1}{\epsilon^3}&\epsilon&-\epsilon^2
\end{matrix}
\right)\left(
\begin{matrix}a_1
\\
a_2
\\
a_3
\end{matrix}
\right).
\end{gather*}
We can not save the representation.
As in the previous case, it is possible to use other models to show that this contraction yields
information about limits of non-Gaussian hypergeometric functions, not related to the Askey scheme.

\section{Nondegenerate to degenerate contractions}\label{Section6}

This appears initially a~mere restriction of the
3-parameter potential to 1-parameter.
However, after restriction one 2nd order generator $L_i$ becomes a~perfect square $L_i=X^2$.
The spectrum of~$L_i$ is nonnegative but that of $X$ can take both positive/negative values.
This results in a~virtual doubling of the support of the measure in the f\/inite case.
Also, the commutator of $X$ and the remaining 2nd order symmetry leads to a~new 2nd order symmetry.
In~\cite{KMSH} we will show how this Casimir follows directly as a~contraction from the expression for~$R^2$.

\subsection[Contraction S9 $\to$ S3]{Contraction S9 $\boldsymbol{\to}$ S3}

{\bf 1) Wilson $\boldsymbol{\to}$ special dual Hahn (1st model).}
The quantum system E3~\eqref{S3algebra} is given in the singular limit from system
${\rm S}9$~\eqref{Wilson} by
\begin{gather}
a_2=a_3=\epsilon\to0,
\qquad
a_1=a_1,\nonumber
\\
X'^2=\lim\limits_{\epsilon\rightarrow0}L_1,
\qquad
L_2'=\lim\limits_{\epsilon\rightarrow0}L_2,
\qquad
L_1'=[X',L_2'].\label{S9toS31}
\end{gather}
The operators in this contraction dif\/fer from those given in Subsection~\ref{degeneratesystems} by
a~cyclic permutation of the coordinates $s_i\rightarrow s_{i+1}$.
Now we investigate how the dif\/ference operator realization of ${\rm S}9$ contracts to irreducible
representations of the ${\rm S}3$ symmetry algebra.
This is more complicated since the original restricted algebra is now contained as a~proper subalgebra of
the contracted algebra.

The contraction~\eqref{S9toS31} is realized in the model by setting $\alpha_2=\alpha_3=-1/2$ and
$\alpha_1=\alpha$ (the subscript is dropped in this model since there is now a~sole $\alpha$).
The restricted operators then become $H'=E'$ with $E'=-4(m+1)(m+\alpha)-(\alpha-1)^2+\frac14$ and
\begin{gather*}
X'^2=-4\tau^*\tau,
\qquad
L_2'=-4t^2+\alpha^2-\frac14.
\end{gather*}
The eigenfunctions for $X'^2$, the Wilson polynomials, become
\begin{gather}\label{Wilson1}
\Phi_{\pm n}\big(t^2\big)={}_4F_{3}\left(
\begin{matrix}
-n,
&
n,
&
-\frac{4m+2\alpha+1}{4}-t,
&
-\frac{4m+2\alpha+1}{4}+t
\\
-m,
&
-m-\alpha,
&
\frac12
\end{matrix}
;1\right).
\end{gather}
Here $n=0,1,\dots,m$ if $m$ is a~nonnegative integer and $n=0,1,\dots$ otherwise.
For f\/inite dimensional representations, the spectrum of $t$ is the set $\{
\frac{\alpha}{2}+\frac14+m-k,\, k=0,1,\dots,m\}$.
Note that the restricted polynomial functions~\eqref{Wilson1} are no longer the correct basis functions for
the contracted superintegrable system.
To see this, we consider the contracted expansion coef\/f\/icients
\begin{gather*}
L_2'f_{n,m}=K(n+1,n)f_{n+1,m}+K(n-1,n)f_{n-1,m}+\left(K(n,n)+\alpha_1^2-\frac14\right)f_{n,m},
\end{gather*}
with
\begin{gather*}
K(n+1,n)=\frac14\left(m-n+\alpha\right)\left(m-n\right),
\qquad
K(n-1,n)=\frac14\left(m+n+\alpha\right)\left(m+n\right),
\\
K(n,n)=\left(\frac12\left(\alpha+\frac12\right)-m\right)^2-K(n+1,n)-K(n-1,n).
\end{gather*}
Indeed $K(n-1,n)$ no longer vanishes for $n=0$, so $f_{0,m}$ is no longer the lowest weight eigenfunction.
Note that $f_{-1,m}=f_{1,m}$ is still a~polynomial in $t^2$. To understand the contraction we set $n=N-M/2$
where $N$ is a~nonnegative integer and $M=2m$.
Then the equations for the $K$'s become
\begin{gather*}
K(N+1,N)=\frac14\left(M-N+\alpha\right)\left(M-N\right),
\qquad
K(N-1,N)=\frac14N\left(N+\alpha\right),
\end{gather*}
and the three term recurrence relation gives a~new set of orthogonal polynomials for representations of
${\rm S}3$.
The lowest eigenfunction occurs for $N=0$; if $M$ is a~nonnegative integer the representation is
$(2m+1)$-dimensional with highest eigenfunction for $N=M$.
The basis functions which satisfy this three-term recurrence are
\begin{gather}\label{dualHahna1}
f_{N,M}\big(t^2\big)=\frac{(\alpha+1)_N}{(-\alpha-M)_N}{}_3F_2\left(
\begin{matrix}
-N,&-s,&s+2\alpha+1
\\
-M,&1+\alpha
\end{matrix}
;1\right).
\end{gather}
The relation between $t$ and $s$ is $s=2t-\alpha-1/2$. Here $f_N$ is a~polynomial of order $2N$ in $s$ and
of order $n$ in $\lambda(s)=s(s+2\alpha+1)$, a~special case of dual Hahn polynomials.
These special dual Hahn polynomials are associated with the dif\/ference operator,
\begin{gather}\label{XS3dualHahn}
X=i\left(B(s)E_s+C(s)E_s^{-1}\right),
\end{gather}
with $B(s)+C(s)=M$ def\/ined as
\begin{gather*}
B(s)=\frac{(s+2\alpha+1)(M-s)}{2s+2\alpha+1},
\qquad
C(s)=\frac{s(s+M+2\alpha+1)}{2s+2\alpha+1}.
\end{gather*}
The operators which form a~model for the algebra~\eqref{S3algebra} are $X$~\eqref{XS3dualHahn} along with
\begin{gather}\label{S3Model1}
L_1=-\left(s+\alpha+\frac12\right)^2+\alpha^2-\frac14,
\qquad
L_2=[L_1,X].
\end{gather}
For f\/inite dimensional representations the spectrum of $s$ is $\{0,1,\dots,M\}$.

What is the relation between the functions~\eqref{Wilson1} and the proper basis
functions~\eqref{dualHahna1}? Note this model $X$, $L_1$, $L_2$ can be obtained from the contracted model $X'$,
$L_1'$, $L_2'$ by conjugating by the ``ground state'' of the contracted model $\Phi_{-\frac{M}{2}}\big(t^2\big)$. We
f\/ind explicitly the gauge function
\begin{gather*}
\Phi_{-\frac{M}{2}}\big(t^2\big)
=\frac{\big(\frac12-\frac{M}{2}\big)_k(-\alpha-M)_k}{\big(\frac12\big)_k\big(-\alpha-\frac{M}{2}\big)_k},
\end{gather*}
when $t$ is evaluated at the weights
\begin{gather}\label{tweights}
t=\frac{\alpha}{2}+\frac14+\frac{M}{2}-k,
\qquad
k=0,1,\dots,\frac{M}{2}.
\end{gather}
So, the operator $X'$ is related to $X$ via conjugation by $\Phi_{-\frac{M}{2}}\big(t^2\big)$.

Note that the functions $\Phi_n\big(t^2\big)$ are only def\/ined for discrete values of $t$~\eqref{tweights}.
However, on this restricted set the functions $\Phi_{-\frac{M}{2}+N}$ and $f_{N,M}$ satisfy exactly the
same three term recurrence formula under multiplication by $-4t^2-a$, with the bottom of the weight ladder
at $N=0$.
From this we f\/ind the identity
\begin{gather*}
\Phi_{-\frac{M}{2}}\big(t^2\big)f_{N,M}\big(t^2\big)=\Phi_{-\frac{M}{2}+N}\big(t^2\big),
\qquad
t=\frac{\alpha}{2}+\frac14+\frac{M}{2}-k.
\end{gather*}
Since $\Phi_{-\frac{M}{2}}\big(t^2\big)=\Phi_{\frac{M}{2}}\big(t^2\big)$
this relation implies that $f_{M,M}\big(t^2\big)=1$ when
restricted to the spectrum of $t$.

{\bf 2) Wilson $\boldsymbol{\to}$ special Hahn (2nd model).}
The quantum system ${\rm S}3$~\eqref{S3Structure} can also be obtained from system ${\rm S}9$~\eqref{Wilson} by
\begin{gather*}
a_1=a_3=\epsilon\to0,
\qquad
a_2=\frac14-\alpha^2,
\\
L_1'=\lim\limits_{\epsilon\rightarrow0}L_1,
\qquad
X'^2=\lim\limits_{\epsilon\rightarrow0}L_2,
\qquad
L_2'=[X',L_1'].
\end{gather*}
Again, the physical model obtained by this contraction is related to the that given in
Subsection~\ref{degeneratesystems} by a~cyclic permutation of the coordinates $s_i\rightarrow s_{i-1}$.

In this limit, the operator $X^2$ can be immediately factorized to obtain the skew-adjoint operator
$X=2it$. Taking $x=t+m$, we f\/ind the operators in our model are
\begin{gather}
L_1'=-\big[B(x)E_x+C(x)E_x^{-1}-B(x)-C(x)\big]-\alpha-\frac12,\nonumber
\\
B(x)=(x-2m)(x+\alpha+1),
\qquad
C(x)=x(x-2m-\alpha-1),\nonumber
\\
X'=2i(x-m),
\qquad
L_2'=[X',L_1],\label{S3Model2}
\end{gather}
which is diagonalized by Hahn polynomials
\begin{gather*}
\widehat{f}_{k,m}={}_3F_2\left(
\begin{matrix}
-k,& k+2\alpha+1, & -x
\\
\alpha+1, & -2m &
\end{matrix}
;1\right)=Q_k(x;\alpha,\alpha,2m),
\\
L_1'\widehat{f}_{k,m}=\left(-\left(k+\alpha+\frac12\right)^2+\alpha^2-\frac14\right)\widehat{f}_{k,m},
\qquad
k=0,1,\dots,2m.
\end{gather*}
These polynomials satisfy special recurrence relations not obeyed by general Hahn polynomials.
Note that the dimension of the representation space has jumped from $m+1$ to $2m+1$. Comparing these
eigenfunctions with the limit of the Wilson polynomials,
\begin{gather*}
\lim\limits_{\epsilon\rightarrow0}L_1'f_{n,m}=\left(-\left(2n+\alpha+\frac12\right)^2+\alpha-\frac14\right)f_{n,m},
\\
f_{n,m}(t)={}_4F_3\left(
\begin{matrix}
-n,&n+\alpha+\frac12,&-m-t,&-m+t
\\
-m,&\frac12-m,&\alpha+1
\end{matrix}
;1\right),
\end{gather*}
$n=0,1,\dots,m$, we see that in the limit only about half of the spectrum is uncovered.
Note that, the functions $f_{n,m}$ are even functions of $t$ whereas
$\widehat{f}_{k,m}(-t)=(-1)^k\widehat{f}_{k,m}(t)$.

The recurrences for multiplication by $2it$ and $-4t^2$ are compatible, so we obtain the following identity
relating a~special case of Wilson polynomials with a~special case of the Hahn polynomials:
\begin{gather*}
{}_4F_3\left(
\begin{matrix}
-n,&n+\alpha+\frac12,&-m-t,&-m+t
\\
-m,&\frac12-m,&\alpha+1
\end{matrix}
;1\right)
\\
\qquad=
{}_3F_2\left(
\begin{matrix}-2n, & 2n+2\alpha+1, & -t-m
\\
\alpha+1, & -2m &
\end{matrix}
;1\right),\qquad n=0,1,\dots,m.
\end{gather*}
(This is a~limit of Singh's $q$-series quadratic transformation,~\cite[p.~89]{GR}.)

\subsection[Contraction E1 $\to$ E6]{Contraction E1 $\boldsymbol{\to}$ E6}

{\bf Jacobi $\boldsymbol{\to}$ Gegenbauer.}
By setting $a_1=0$, $a_3=0$ in system E1 we obtain the system E6.
The contraction of the operators takes the form
\begin{gather}\label{E1toE6}
X'^2=\lim\limits_{\epsilon\to0}L_1,
\qquad
L_1'=\lim\limits_{\epsilon\to0}L_2,
\qquad
L_2'=\frac12[L_1',X'].
\end{gather}

To investigate the ef\/fect of this contraction on the models, we begin with the dif\/ferential operator
model of E1~\eqref{E1Model3} with $\beta_1=1/2$ and $\beta_3=1/2$.
As in the previous section, we now have only one parameter so we drop the subscript and simply write
$a_2=1/4-\beta^2$.
After the change of basis~\eqref{E1toE6}, the operator $X'^2$ is then given by $X'^2=E'(x+1)/2$, suggesting
the change of variables $x=2t^2-1$.
The model becomes $H'=E'$ and
\begin{gather}\label{E6model}
X'=\sqrt{E'}t,
\qquad
L_1'=\big(1-t^2\big)\partial_t^2-2t(1+\beta)\partial_t-\beta-\frac12,
\qquad
L_2'=\frac12[L_1',X'].
\end{gather}
The eigenfunctions of $L_1'$ are given by the Gegenbauer polynomials,
\begin{gather*}
C_k^{\beta+1/2}(t)=\frac{(2\beta+1)_k}{k!}\, {}_2F_1\left(
\begin{matrix}-k, & 2\beta+1+k
\\
\beta+1 &
\end{matrix}
; \frac{1-t}{2}\right).
\end{gather*}

The eigenfunction of $L_2$ in the model for E1, contract to eigenfunctions of $L_1'$
\begin{gather*}
f_n={}_2F_1\left(
\begin{matrix}
-n,&\beta+n+1/2
\\
\beta+1&
\end{matrix}
;t^2-1\right).
\end{gather*}
The expansion coef\/f\/icients of the action of $X'^2$ on this basis are
\begin{gather*}
K'(n+1,n)=\frac{2E'(2\beta+2n+1)(\beta+n+1)}{(2\beta+4n+1)(2\beta+4n+3)},
\\
K'(n-1,n)=\frac{E'2n(2n-1)}{(2\beta+4n-1)(2\beta+4n+1)},
\\
K'(n,n)=E'-K'(n+1,n)-K'(n-1,n),
\end{gather*}
which suggests half-integer values for $n$ in the model.
Indeed, the contraction limit of the basis polynomials gives only half the spectrum.
The full spectrum is obtained from eigenfunctions of~$L_1'$ obtained directly as
\begin{gather}\label{E6basis}
g_k(t)=\frac{k!}{(2\beta+1)_k}C_k^{\beta+1/2}(t),
\qquad
k=0,1,2,\ldots.
\end{gather}
These polynomials are related to $f_n$ (the contracted basis) as $f_{n}(t)=g_{2k}(t)$, giving the
following identity:
\begin{gather*}
{}_2F_1\left(
\begin{matrix}
-k,&\beta+k+1/2
\\
\beta+1&
\end{matrix}
;t^2-1\right)={}_2F_1\left(
\begin{matrix}
-2k,& 2\beta+1+2k
\\
\beta+1 &
\end{matrix}
;\frac{t-1}{2}\right).
\end{gather*}

\subsection[Contraction E8$(a_1=0)\to$ E14]{Contraction E8$\boldsymbol{(a_1=0)\to}$ E14}

This contraction leads to Bessel functions, not to orthogonal polynomials.

\section{Degenerate/singular system contractions}\label{Section7}

\subsection[Contraction S3 $\to$ E3]{Contraction S3 $\boldsymbol{\to}$ E3}

{\bf Special dual Hahn $\to$ Special Krawtchouk.}
In model~\eqref{S3Model1} we set $\tilde{a}=\epsilon^2a$,
\begin{gather*}
\left(
\begin{matrix}{\tilde L_1}
\\
{\tilde L_2}
\\
{\tilde H}
\\
{\tilde X}
\end{matrix}
\right)=\left(
\begin{matrix}
\epsilon&0&0&0
\\
0&\epsilon&0&0
\\
0&0&\epsilon&0
\\
0&0&0&1
\end{matrix}
\right)\left(
\begin{matrix}L_1
\\
L_2
\\
H
\\
X
\end{matrix}
\right)+\left(
\begin{matrix}0
\\
0
\\
-\epsilon
\\
0
\end{matrix}
\right)a
\end{gather*}
and obtain in the limit $L_i'=\lim\limits_{\epsilon\rightarrow}\tilde{L}_i$ with $a'=\lim\limits_{\epsilon\rightarrow
}\tilde{a}\equiv-\omega^2$.
The model becomes
\begin{gather*}
H'=-2\omega(M+1),
\qquad
iX'=(s-M)E^1-sE^{-1},
\\
L_1'=\omega(2s+2M+3),
\qquad
L_2'=-\omega(s-M)E^{1}-sE^{-1}.
\end{gather*}
The basis functions for this representation are
\begin{gather*}
f_N=(-1)^N{}_2F_1\left(
\begin{matrix}
-N,&-s
\\
-M
\end{matrix}
;2\right),
\end{gather*}
special Krawtchouk or Meixner polynomials, depending on whether $M$ is a~positive integer.
The eigenvalues of $iX'$ are $M,M-2,\dots,-M$ in the f\/inite case.
The action of $L'_1$ is $L_1'f_N=(N-M)f_{N+1}-(M+1)f_N-Nf_{N-1}$, and the action of $L_2'$ follows from
commutation relation $[L_1',X']=2L_2'$.

For the second model~\eqref{S3Model2}, after contraction we have
\begin{gather*}
L_1=\omega\left((x-2m)E+xE^{-1}(-2m-1)\right),
\qquad
X=2i(x-m),
\\
L_2=-i\omega\left((x-2m)E-xE^{-1}\right).
\end{gather*}
The eigenvalues of $L_1$ are $-\omega(2k+1)$, $k=0,1,\dots,2m$ for f\/inite dimensional representations,
and the corresponding eigenfunctions are special:
\begin{gather*}
f_k(x)={}_2F_1\left(
\begin{matrix}
-k,&-x
\\
-2m
\end{matrix}
;2\right).
\end{gather*}

\subsection[Contraction E1$(a_1=0)\to{\rm sl(2)}$]{Contraction E1$\boldsymbol{(a_1=0)\to{\rm sl(2)}}$}

{\bf Jacobi $\boldsymbol{\to}$ Laguerre.} We use model~\eqref{E1Model3} and let
$\beta_3=1/\sqrt{\epsilon}$, $E'=2/\epsilon$, $(1-x)/2=\epsilon v$, $\epsilon\to 0$.
The new basis functions are
\begin{gather}
g_{n}={}_1F_1\left(
\begin{matrix}
-n
\\
\beta_2+1
\end{matrix}
;v\right),
\label{basissl2}
\end{gather}
Laguerre polynomials.
The operators that correspond to this limit are
\begin{gather*}
S_1=\lim\limits_{\epsilon\rightarrow0}H-L_1,
\qquad
S_2=\lim\limits_{\epsilon\rightarrow0}\epsilon L_2,
\qquad
K=\lim\limits_{\epsilon\rightarrow0}L_1.
\end{gather*}
The model contracts to $K=1$ and
\begin{gather*}
S_1=2v,
\qquad
S_2=4v\partial_{vv}+4(-v+\beta_2+1)\partial_v-2(\beta_2+1),
\end{gather*}
whose action on the basis~\eqref{basissl2} is
\begin{gather*}
\begin{split}
& S_1g_{n}=2vg_n=-2(\beta_2+n+1)g_{n+1}-2(\beta_2+2n+1)g_n-2ng_{n-1},
\\
& S_2g_n=\big[4v\partial_{vv}+4(-v+\beta_2+1)\partial_v-2(\beta_2+1)\big]g_{n}.
\end{split}
\end{gather*}
The corresponding limit in the physical system is obtained by taking $y=y'/\sqrt{\epsilon}$,
$\beta_3=1/\epsilon$, so that the limit corresponds to a~subclass of singular quantum systems.
Indeed, letting $\epsilon\to 0$, $K=\frac{1}{y'^2}$ (a constant that we can set to $1$),
$S_1=\partial_{xx}+\frac{\frac14-\beta_2^2}{x^2}$, $S_2=\partial_{xx}+\frac{\frac14-\beta_2^2}{x^2}-x^2$,
we f\/ind that $S_1$, $S_2$ generate the Lie structure algebra sl(2).

\subsection[Contraction E6 $\to$ oscillator algebra]{Contraction E6 $\boldsymbol{\to}$ oscillator algebra}

{\bf Gegenbauer $\boldsymbol{\to}$ Hermite.} The E6 algebra contracts to a~Lie algebra under the following limit
\begin{gather}\label{E6toOscillator}
\hat{L}_1=\lim\limits_{\epsilon\rightarrow0}\epsilon L_1,
\qquad
\hat{L}_2=\lim\limits_{\epsilon\rightarrow0}\epsilon L_2,
\qquad
\hat{H}=\lim\limits_{\epsilon\rightarrow0}\epsilon H,
\qquad
\hat{X}=X,
\qquad
a=\frac{1}{\epsilon^2}.
\end{gather}
Under this contraction, the algebra relations become
\begin{gather*}
[\hat{L}_2,\hat{X}]=2\hat{L}_1,
\qquad
[\hat{L}_2,\hat{L}_1]=2\hat{X},
\qquad
[\hat{X},\hat{L}_1]=-\hat{H}.
\end{gather*}

We use the Gegenbauer model~\eqref{E6model} for E6 and with basis functions $g_k(t)$~\eqref{E6basis}.
The representation can be saved by taking $\beta=1/\epsilon$, $t=\sqrt{\epsilon}u$ and
$E=\hat{E}/\epsilon$.
The model becomes
\begin{gather*}
\hat{X}=\sqrt{\hat{E}}u,
\qquad
\hat{L}_1=\partial_{uu}-2u\partial_u-1,
\qquad
\hat{L}_2=\frac12[\hat{L}_1,\hat{X}].
\end{gather*}
In this limit, the Gegenbauer polynomials tend to the Hermite polynomials,
\begin{gather*}
\frac{H_k(u)}{2^{k/2}k!}=\lim\limits_{\lambda\to\infty}\frac{C^\lambda_k(\frac{u}{\sqrt{\lambda}})}{\lambda^{k/2}},
\qquad
\lambda=\beta+\frac12\rightarrow\infty.
\end{gather*}

In the original quantum system we set $x=t/\sqrt{\epsilon}$ and $a=1/\epsilon^2$.
Then, for $\epsilon\to 0$ in the contraction~\eqref{E6toOscillator}, we have
\begin{gather*}
L_1=t^2\partial_{yy}-y^2/t^2,
\qquad
L_2=y/t^2,
\qquad
X=\partial_y,
\qquad
H=-1/t^2.
\end{gather*}
Note that operators $X$, $L_1$, $L_2$, $H$ determine the oscillator algebra, the Lie algebra generated by the
annihilation/creation operators for bosons, $a$, $a^*$, the number of particles operator $N=a^*a$ and the
identity $I$.

\section{Contractions to Laplace--Beltrami equations}\label{Section8}

\subsection[Contraction E6 $\to$ e(2)]{Contraction E6 $\boldsymbol{\to}$ e(2)}

{\bf Jacobi $\boldsymbol{\to}$ Tchebichef\/f.} We contract to the free space Hamiltonian by setting
$a \to 0$, i.e.\
$\beta \to -1/2$. In the limit we f\/ind the continuous Tchebichef\/f polynomials
$g_k(x)=2T_k(x)=k\lim\limits_{s\to 0}\frac{C_k^s(v)}{s}$.

After the contraction we have ${\hat L_2}={\hat J}^2$, ${\hat J}=\sqrt{1-v^2}\partial_v$, and ${\hat J}$
generates the new symmetry ${\hat X_1}=[{\hat J},{\hat X_2}]=2iM\sqrt{1-v^2}$.
Since $[{\hat X_1},{\hat X_2}]=0$ and ${\hat L}_1=\frac12\{{\hat J},{\hat X_1}\}$, the quadratic algebra
now closes to e(2) the Euclidean Lie algebra.

\subsection[Contraction ${\rm S}3\to{\rm so(3)}$]{Contraction $\boldsymbol{{\rm S}3\to{\rm so(3)}}$}

{\bf Dual Hahn $\to$ Special Krawtchouk.}
We use model~\eqref{dualHahna1}.
In the f\/inite case the multiplicity of $E$ is $M+1$.
To go to the Laplace--Beltrami eigenvalue equation on the sphere we let $\alpha\to -1/2$.
In this limit, cancellation occurs and we have $X=\frac{(s-M)}{2}E^1-\frac{(s+M)}{2}E^{-1}$, $L_2=S^2$,
$S=is$.
From $[X,S]$ we obtain $Y=\frac{(s-M)}{2}E^1+\frac{(s+M)}{2}E^{-1}$ and these 3 generators def\/ine the Lie
algebra so(3).
With this new symmetry the dimension of the f\/inite representations becomes $2M+1$, the basis functions
are polynomials in $s$, rather than $s^2$ and the spectrum of $s$ is $-M,-M+1,\dots,M$.
The new basis polynomials are
\begin{gather*}
{}_2F_1\left(
\begin{matrix}
-N,&-s-M
\\
-2M
\end{matrix}
;2\right),
\end{gather*}
special Krawtchouk polynomials $K_N(s+M;\frac12,2M)$ in the f\/inite dimensional case and special Meixner
polynomials $M_N(s+M;-2M,-1)$ in the inf\/inite case.

\begin{figure}[t]
\centering
\includegraphics[width=12cm]{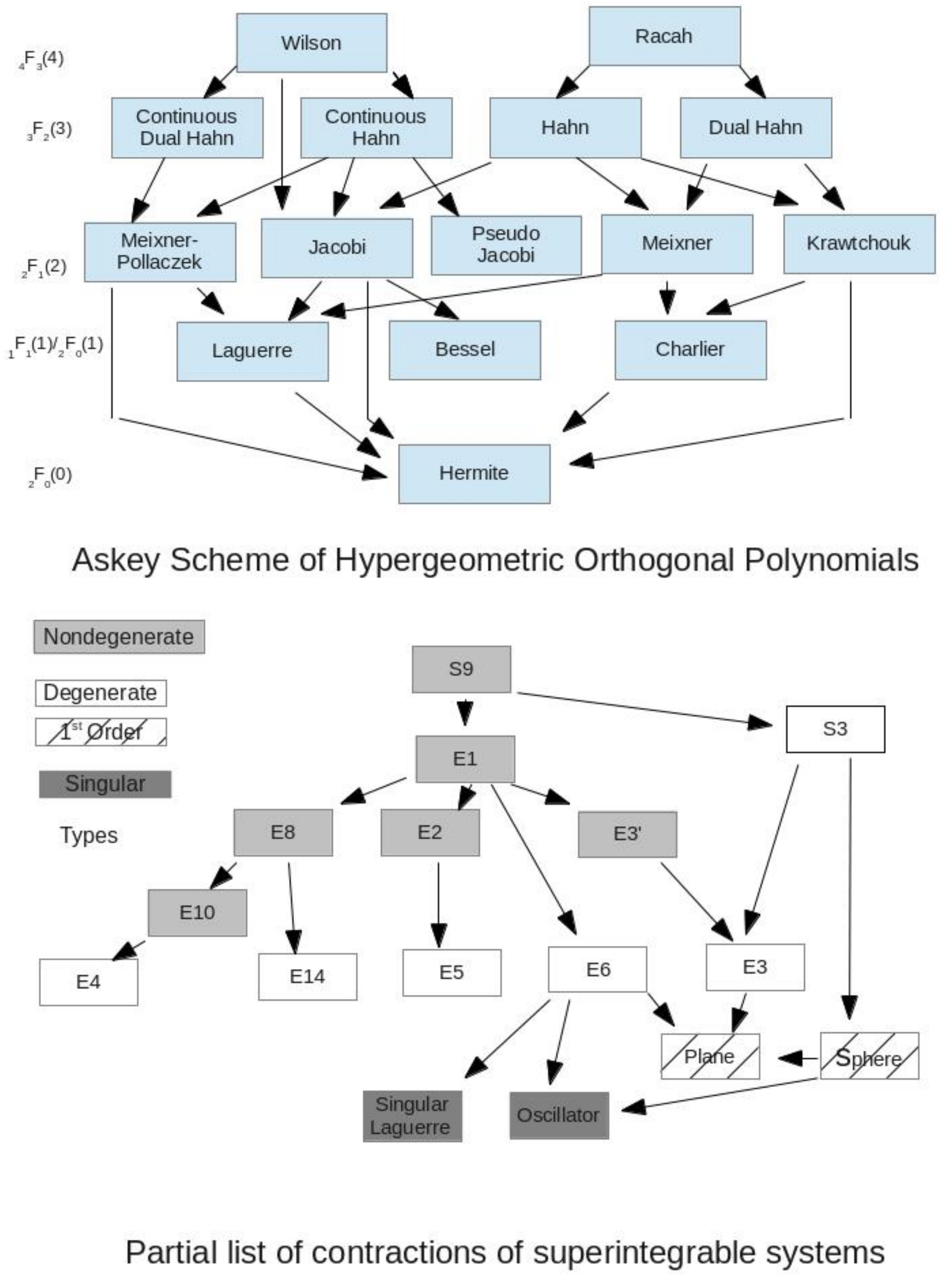}
\caption{The Askey scheme and contractions of superintegrable systems.}
\label{fig1}
\end{figure}

\begin{figure}[t]
\centering
\includegraphics[width=12cm]{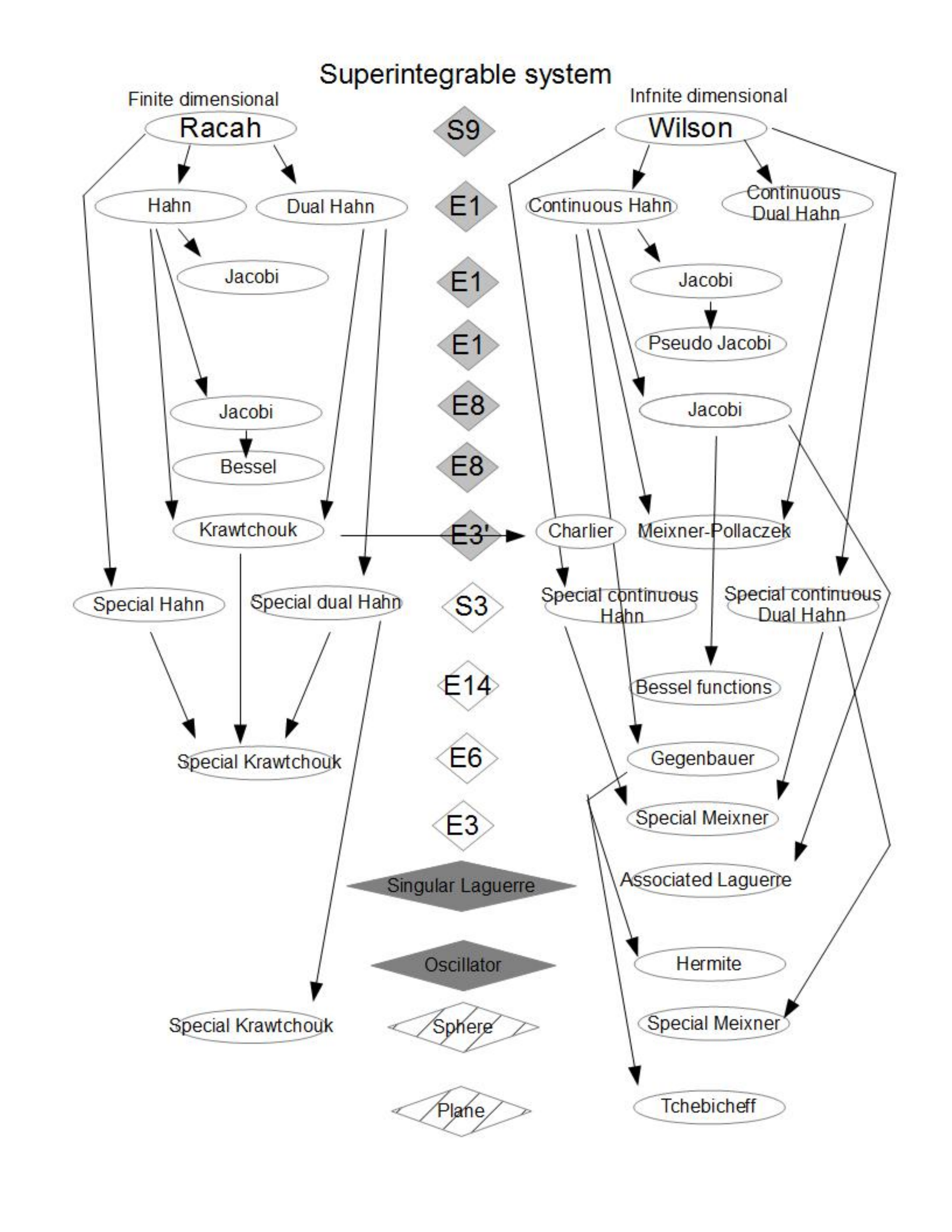}
\caption{The Askey contraction scheme.}
\label{fig2}
\end{figure}

\section{The contraction scheme and f\/inal comments}\label{Section9}

The top half of Fig.~\ref{fig1} shows the standard Askey scheme indicating which orthogonal polynomials can
be obtained by pointwise limits from other polynomials and, ultimately, from the Wilson or Racah
polynomials.
The bottom half of Fig.~\ref{fig1} shows how each of the superintegrable systems can be obtained by
a~series of contractions from the generic system ${\rm S}9$.
Not all possible contractions are listed, partly due to complexity and partly to keep the graph from being
too cluttered.
(For example, {\it all} nondegenerate and degenerate superintegrable systems contract to the Euclidean
system $H=\partial_{xx} +\partial_{yy}$.) The {\it singular systems} are superintegrable in the sense that
they have 3 algebraically independent generators, but the coef\/f\/icient matrix of the 2nd order terms in
the Hamiltonian is singular.
Fig.~\ref{fig2} shows which orthogonal polynomials are associated with models of which quantum
superintegrable system and how contractions enable us to reach all of these functions from ${\rm S}9$.
Again not all contractions have been exhibited, but enough to demonstrate that the Askey scheme is
a~consequence of the contraction structure linking 2nd order quantum superintegrable systems in~2D.
It is worth remarking that forthcoming papers by us will simplify considerably the compexity of our
approach,~\cite{KMSH}.
We will show that the structure equations for nondegenerate superintegrable systems can be derived directly
from the expression for $R^2$ alone, and the structure equations for degenerate superintegrable systems can
be derived, up to a~multiplicative factor, from the Casimir alone.
It will also be demonstrated that all of the contractions of quadratic algebras in the Askey scheme can be
induced by natural contractions of the Lie algebras ${\rm e}(2,{\mathbb C})$ (6~possible contractions) and
${\rm o}(3,{\mathbb C})$ (4 possible contractions).

This method obviously extends to 2nd order systems in more variables.
A start on this study can be found in~\cite{KMPost11}.
To extend the method to Askey--Wilson polynomials we would need to f\/ind appropriate $q$-quantum
mechanical systems with $q$-symmetry algebras and we have not yet been able to do so.

\subsection*{Acknowledgment} This work was partially supported by a~grant from the Simons Foundation (\#~208754 to Willard Miller, Jr.).
The authors would also like to thank the referees for their valuable comments and suggestions.

\pdfbookmark[1]{References}{ref}
\LastPageEnding

\end{document}